\documentclass[preprint]{aastex}
\usepackage{float}
\usepackage{placeins}

\newcommand{\fe}{[Fe/H]}
\newcommand{\kms} {km s$^{-1}$}

\slugcomment{revised version, 2015 Nov 10}

\shorttitle{Quantifying Galactic Halo Substructure}
\shortauthors{Janesh et al.}

\begin{document}

\title{The SEGUE K Giant Survey. III. Quantifying Galactic Halo Substructure}
\author{William Janesh\altaffilmark{1,*}, Heather L. Morrison\altaffilmark{1}, Zhibo Ma\altaffilmark{1},  Constance Rockosi\altaffilmark{2}, Else Starkenburg\altaffilmark{3,4,5}, Xiang Xiang Xue\altaffilmark{6}, Hans-Walter Rix\altaffilmark{6}, Paul Harding\altaffilmark{1}, Timothy C. Beers\altaffilmark{7}, Jennifer Johnson\altaffilmark{8,9}, Young Sun Lee\altaffilmark{10}, Donald P. Schneider\altaffilmark{11,12}}

\altaffiltext{1}{
Department of Astronomy, Case Western Reserve University, Cleveland, OH 44106, USA
}
\altaffiltext{*}{
Current Address: Department of Astronomy, Indiana University, Bloomington, IN 47405, USA
}
\altaffiltext{2}{
UCO/Lick Observatory, University of California, Santa Cruz, 1156 High Street, Santa Cruz, CA 95064, USA
}
\altaffiltext{3}{
Department of Physics and Astronomy, University of Victoria, PO Box 1700, STN CSC, Victoria BC V8W 3P6, Canada
}
\altaffiltext{4}{
CIFAR Global Scholar
}
\altaffiltext{5}{
Leibniz-Institut f\"{u}r Astrophysik Potsdam, An der Sternwarte 16, 14482 Potsdam, Germany
}
\altaffiltext{6}{
Max-Planck-Institut f\"{u}r Astronomie, K\"{o}nigstuhl 17, D-69117 Heidelberg, Germany
}
\altaffiltext{7}{
Department of Physics and JINA Center for the Evolution of the Elements, University of Notre Dame, Notre Dame, IN 46556, USA
}
\altaffiltext{8}{
Department of Astronomy, Ohio State University, 140 West 18th Avenue, Columbus, OH 43210, USA
}
\altaffiltext{9}{
Center for Cosmology and Astro-Particle Physics, Ohio State University, Columbus, OH 43210, USA
}
\altaffiltext{10}{
Department of Astronomy and Space Science, Chungnam National University, Daejeon 34134, Republic of Korea
}
\altaffiltext{11}{
Department of Astronomy and Astrophysics, The Pennsylvania State University, University Park, PA 16802
}
\altaffiltext{12}{
Institute for Gravitation and the Cosmos, The Pennsylvania State University, University Park, PA 16802
}

\begin{abstract}
We statistically quantify the amount of substructure in the Milky Way stellar halo using a sample of 4568 halo K giant stars at Galactocentric distances ranging over 5-125 kpc. These stars have been selected photometrically and confirmed spectroscopically as K giants from the Sloan Digital Sky Survey's SEGUE project. Using a position-velocity clustering estimator (the 4distance) and a model of a smooth stellar halo, we quantify the amount of substructure in the halo, divided by distance and metallicity. Overall, we find that the halo as a whole is highly structured. We also confirm earlier work using BHB stars which showed that there is an increasing amount of substructure with increasing Galactocentric radius, and additionally find that the amount of substructure in the halo increases with increasing metallicity. Comparing to resampled BHB stars, we find that K giants and BHBs have similar amounts of substructure over equivalent ranges of Galactocentric radius. Using a friends-of-friends algorithm to identify members of individual groups, we find that a large fraction ($\sim$33\%) of grouped stars are associated with Sgr, and identify stars belonging to other halo star streams: the Orphan Stream, the Cetus Polar Stream, and others, including previously unknown substructures. A large fraction of sample K giants (more than 50\%) are not grouped into any substructure. We find also that the Sgr stream strongly dominates groups in the outer halo for all except the most metal-poor stars, and suggest that this is the source of the increase of substructure with Galactocentric radius and metallicity.

\end{abstract}

\defcitealias{starkenburg}{S09}
\defcitealias{lm10}{LM10}

\section{Introduction}
Current cosmological models predict that structure forms through hierarchical processes. For galaxies, the hierarchical assembly model implies that satellite galaxies will be tidally disrupted, leaving stellar debris \citep{bullock05, cooper10}. Large--scale surveys like the Sloan Digital Sky Survey \citep[SDSS;][]{sdss} and the Two Micron All Sky Survey \citep[2MASS;][]{2mass} have the capability to find these substructures around the Milky Way. Kinematic and chemical information derived from these observations then allow for more detailed analysis of the history of the buildup of the Galaxy.

The search for stellar substructure in and around the Milky Way has been remarkably successful \citep[cf.][]{belokurov}. In addition to the the Sagittarius dwarf spheroidal galaxy \citep[Sgr;][]{ibatasgr}, and its associated tidal stream \citep{mateo96,majewski03}, about a dozen presumably distinct streams or overdensities have been discovered photometrically \citep{duffau06, gd06, belokurov07, newberg07, newbergcetus, newbergorph, grillmairhh}. These streams have a variety of morphologies, ranging from thin, kinematically cold streams to `clouds', and may comprise the majority of all halo stars beyond $\sim 15$ kpc \citep{bell08}.

% are examples of massive stellar objects being accreted in the stellar halo. Other substructure of more uncertain origin exists in the halo as well. The Monoceros stream is a relatively wide stream nearby the Milky Way's thick disk \citep{belokurov}. The Orphan stream is a cold, narrow stream of unknown origin, but possessing a metal-weak composition \citep{newbergorph,belokurov07}. The Cetus Polar Stream, on a polar orbit of the Milky Way, is a fairly distant stream which also has a low metallicity \citep{newbergcetus}. The constellation of Virgo is a substructure goldmine, possessing no fewer than three kinematically different streams or overdensities \citep{duffau06,newberg07}. The Grillmair-Dionatos stream is another nearby cold, narrow stream without a known progenitor \citep{gd06}. This admittedly incomplete list shows the wide variety of substructure that has been discovered, but there have been relatively few attempts to quantify its contribution to the Milky Way halo.

By collecting the spatial and kinematic properties of Galactic stars, a statistical analysis of substructure in the Galactic stellar halo becomes possible. Precedent for this kind of analysis is widespread. Maps of stellar density are a classical method of finding substructure. Substructure regions will appear to be more dense on the sky than a smooth background of stars. Large--scale photometric surveys are well suited to this method. For example, \citet{belokurov} used a simple color cut on a sample of SDSS stars which highlighted a number of structures in the Milky Way halo, including the Sgr tidal streams, the Orphan Stream, and Monoceros ring. With multi-color photometric data, this analysis can also be extended to population information. %\citet{bell10} used SDSS data to determine the ratio of blue horizontal branch (BHB) stars to main-sequence-turnoff stars and found that this ratio varied between different structures, suggesting a variation of properties between and even internal to progenitors.
\citet{bell08} quantified the variation in star counts in the color
region dominated by halo main sequence stars, and \citet{alys11} made
a similar calculation using blue horizontal branch (BHB hereafter)
stars, and found a smaller variation in star counts at the same
distances. This result is intriguing because, while all stars start
their lives on the main sequence, only those with a certain range of
age and metallicity evolve to become BHB stars, which are typical of
old, metal poor populations. Thus the smaller amount of substructure
seen in BHB stars may be due to a population difference rather than
simply indicating a smoother halo than previously thought, as claimed
by  \citet{alys11}.

%Great circle counts \citep{johnston96} are a variation on this method which use star counts in great circles to determine the pole of a stream's orbit. Orbital poles are also used in \citet{lb95}'s method, which attempts to use the orbital energies and angular momenta of globular clusters in streams to identify possible progenitors. Although successful in confirming well-studied globular clusters to belong in certain streams, this method is quite subjective in its interpretation. 

Use of kinematic data enhances a metric's ability to distinguish substructure from its surrounding smooth distribution. \citet{gorski89}'s method is similar to the classic two-point correlation function, in that it uses a random smooth distribution for significance testing, but the metric instead finds the probability that two objects separated by a given distance will have the same velocity, rather than the number of pairs at a given separation in distance. This metric has been largely used in cosmological clustering studies, but could be easily adapted to stellar kinematic data. \cite{schlaufman} identify substructure by exploring the radial velocity distributions of metal-poor main-sequence turnoff stars on individual SDSS plug-plates (hereafter, plates). This tracer limits them to the inner halo: distances of 17.5 kpc and less. This method requires a fairly dense sample to be effective, since it makes use of velocity distributions. \cite{schlaufman} used nearby SDSS dwarfs in their analysis, but were still able to identify ten stellar structures in the nearby halo and make predictions about the overall membership of stars in substructure, finding that $1/3$ of the volume of the local halo contains velocity substructure.

The logical extension of this progression is to use all six dimensions of phase space $(x,y,z,v_x,v_y,v_z)$ to identify substructure. Given the very accurate observations that are required to measure all six of these dimensions, and the current capabilities of large--scale ground based surveys, only four of these dimensions are readily available. \citet{starkenburg} used a sample of 101 K giants from the Spaghetti survey \citep{spag}, with distances up to $\sim$100 kpc and line-of-sight velocities to determine the overall amount of substructure in the Milky Way stellar halo, finding that it is indeed highly structured, but less so than simulated halos derived from \citet{harding01}. This difference is due to the sparse spatial sampling of the Spaghetti survey being unable to resolve narrow streams in the halo, an issue which \citet{starkenburg} suggest could be alleviated by a much larger survey. Additionally, the authors were able to identify a number of stars associated with the Sagittarius stream and Virgo overdensity using a friends-of-friends group finding algorithm. 

Analysis of a much larger sample of SDSS BHB stars with $5~\textrm{kpc} < R_{gc} < 60~\textrm{kpc}$ by \citet{xue11} (hereafter X11) demonstrated with very high significance that position--velocity substructure exists throughout the Milky Way halo. With this sample, X11 showed that the outer halo was more structured than the inner halo, and that these observations were consistent with, but less prominent than, levels of substructure in mock catalogs derived from the \citet{bullock05} models. X11 attributed
the difference with the models to the strong representation of BHB
stars in older stellar populations, showing that the older model
particles also exhibited less substructure. Further, \citet{cooper11} used ($x,y,z$) and line-of-sight velocity to quantify substructure in mock catalogs from the Aquarius stellar halo models \citep{aquarius,cooper10}, comparing this with a smaller
sample of SDSS BHB stars from \citet{xue08}. Interestingly, they find
that the clustering signal for stars 20-60 kpc from the Galactic
center is consistent with the range of clustering in the Aquarius
models, but for inner halo stars the signal is lower than {\it all} of
the model halos. They suggest that this lack of substructure may be due
to a smooth component of the inner halo.

Studying the amount of halo substructure in red giants is preferable to solely using BHB stars, as all intermediate-age and old stars, regardless of metallicity, turn into red giants. We note that K giants are drawn from a larger range of age and metallicity than BHB stars, and so comparisons between substructure in the two groups could be very helpful. In this study, we present a considerably larger set of red giants than in \citet{starkenburg} to overcome the problems of statistical sampling and provide a more robust comparison with simulations. We use a statistical method called the 4distance \citep[][described fully in Section \ref{4d}]{starkenburg} to quantify the substructure in the largest and most spatially diverse spectroscopic K giant sample to date. In Section \ref{kg}, we describe our sample and the methods used to construct it. In Section \ref{4d} we also discuss other methods used in this paper, including normalization techniques and the friends-of-friends group finding algorithm. In Sections \ref{4dresults} and \ref{fofresults}, we present our results.

\section{The sample}\label{kg}

The Sloan Extension for Galactic Understanding and Exploration (SEGUE) obtained nearly 350,000 spectra of 21 different target types in its four years of observations (\citealp{segue}, C. Rockosi et al. in preparation, \citealp{sdss,2006AJ....131.2332G,2011AJ....142...72E,2013AJ....146...32S}). The definition, calibration and verification of our K giant sample is described in detail in H. Morrison et al. (in preparation), and the Bayesian distance estimation technique we employ is described in \citet{xue14}. Here we provide an overview only. 

SEGUE used a pencil-beam sampling of the sky, which can be seen in Figure~\ref{footprint}. The SDSS $ugriz$ system \citep{ugriz} was not designed for studying stars, so targeting possible giants was not as easy as it would have been with, for example, specially designed filters such as the DDO 51 filter \citep{doug84}. Three target types aimed to identify K giants; all three used regions of the $u-g$/$g-r$ color-color diagram. The first target type ($l$-color K giants), designed to identify metal-poor giants of the halo, used the metallicity sensitivity of the $u-g$ color at the bluer end of the K giant region to identify halo giants. However, this method does not work for the redder stars: as discussed in \citet{segue} and illustrated in their Figure 10, the giant sequence crosses the foreground dwarf locus around $(g-r)_0$=0.8, and appears above the foreground locus for even redder $g-r$. SEGUE used two target types to identify K giants here: the red K giant and proper motion K giant categories.

We identified our K giants by first finding all the stars in SDSS data release 9 \citep[DR9;][]{sdssdr9} with spectra which also satisfied the $ugr$ and other requirements for target selection of the three types discussed above. We limited our consideration to stars with $E(B-V)$ from \citet{sfd} less than 0.25 mag. We then used the SEGUE spectra to measure each star's luminosity by first taking a log $g$ value from the SEGUE Stellar Parameters Pipeline \citep[SSPP;][]{sspp} as an initial cut (log $g$ $<$ 3.5), then using the Mg index, mirroring the method described in \citet{morrison03}, to identify giants. As well as the three target selection categories, there were spectroscopically confirmed giants which do not satisfy the selection criteria of any of the three target types described above: we refer to them as serendipitous K giants.

% disk contamination section with Beers & Schneider comments included

% \subsection{Disk contamination \label{nodisk}}
Our overall aim in this paper is to quantify the amount of
substructure in the Milky Way's stellar halo, and thus we need to exclude
(thick) disk stars from our sample.  While SEGUE's giant
classification was aimed at identifying giants in the halo, there are
fields (particularly at low latitude) where thick disk giants
dominate.

We have chosen not to use a model of the structure of the
disk/thick disk in our work, since the structure of the thick disk
away from the solar cylinder, in particular, is still an active area
of research. Instead, we used both spatial cuts and also a kinematic
cut, excluding stars which satisfy {\it all} of the following
criteria: \fe $> -1.0$,$|z| <10$ kpc, the distance from the Galactic
center in the plane $R <20$ kpc, and the star is located inside the
region of the longitude-velocity plot shown in blue in Figure
\ref{showcuts}.

We chose a metallicity cut as disk stars are on average much more
metal-rich than halo stars. Since there are a few halo stars with
[Fe/H] greater than --1.0 even in the solar neighborhood \citep[see,
for example][]{carney96}, and there may be some thick disk stars with
\fe\ less than --1.0 (the metal-weak thick disk stars discussed
below), we used a cut of \fe~ $>-1.0$ as a good compromise.

Making a cut on distance from the center of the Galaxy also required a
compromise. Many Sgr orbit models \citep[e.g.][; LM10 hereafter]{lm10} show perigalactic
distances around 15-20 kpc. By contrast, although the position of the
edge of the Galactic disk is not well constrained, it is likely to be
between 15 and 20 kpc from the Galactic
center \citep{reyle09,carraro10}.  We have chosen 20 kpc as the best
distance limit here.  Our choice of a cut in $z$ height ($|z|<10$ kpc)
might seem high for a thick disk scale height of order 1
kpc \citep{srm93,juric08}. However, given the expectation that the numbers
of a stellar population in a pencil-beam survey peak at 2-3 times its
scale height due to the volume element growing with distance, we
have chosen a conservative cut in order to exclude as many thick disk stars as possible.

\begin{figure}[t!]
\centering
\includegraphics[scale=0.6]{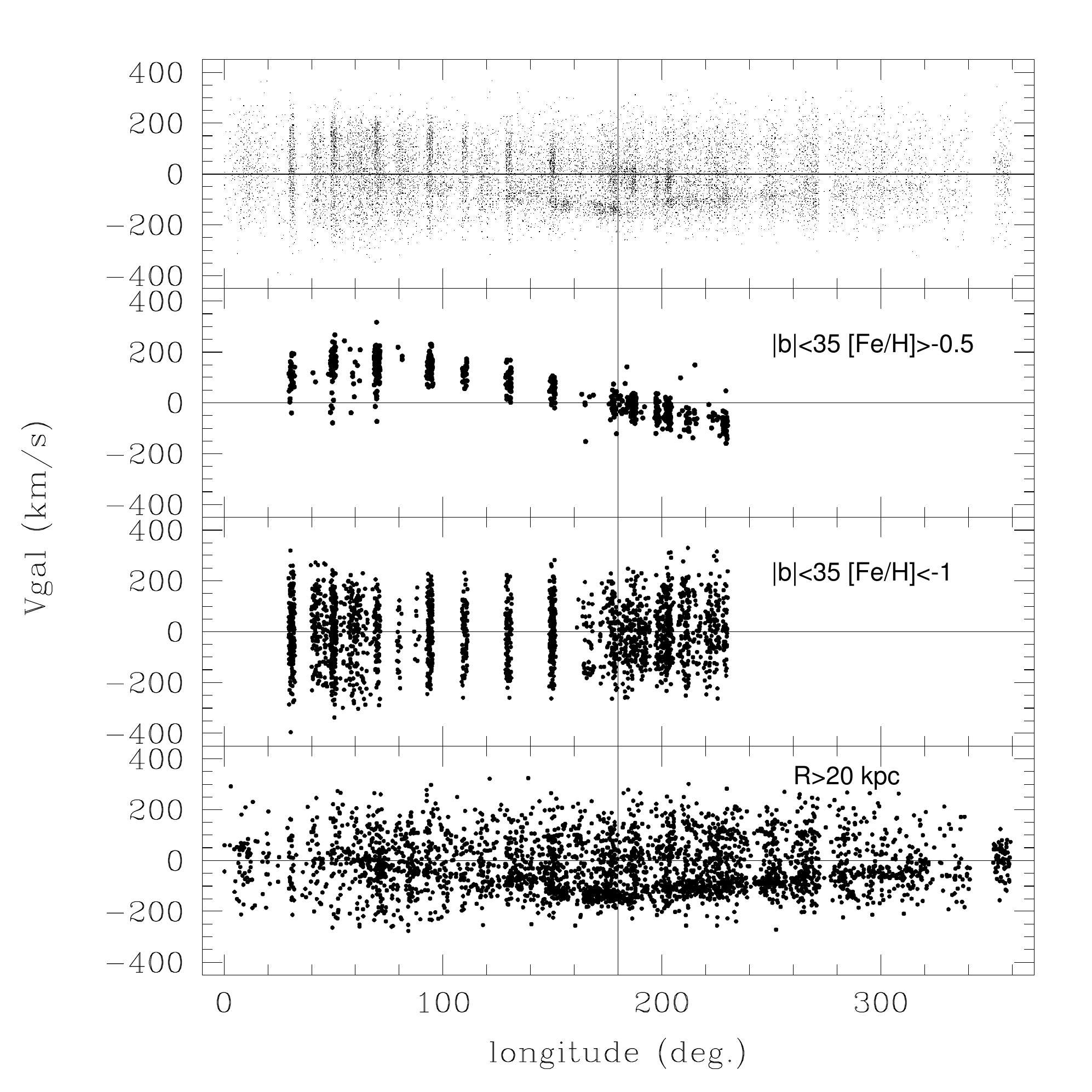}
\caption{The signature of the disk and of the Sgr 
dSph stream in a plot of Galactic longitude $l$ against the line of 
sight velocity of the star, corrected for the projection of the LSR 
velocity along the line-of-sight. The top three panels are drawn from our
extended K giant sample of 15,081 giants, while the bottom panel comes 
from the smaller sample of K giants with distances from X14. 
The top panel shows the entire extended K
giant sample. The next 
lower panel shows stars at low latitude ($|b|<35^\circ$) with high
metallicity (\fe$>-0.5$). The signature of disk rotation is clear here 
(note that we lack stars at low latitude and longitude greater than
 240 deg. because the SDSS telescope is in the North). 
The third panel shows stars with the same low latitude cut, but 
[Fe/H]$<-1.0$, and in contrast, displays a symmetry between negative 
and positive velocities typical of the halo. The last panel includes 
giants with distances greater than 20 kpc from the Galactic center. The
signature of the Sgr dSph stream is clearly seen here.}
\label{nodistlv}
\end{figure}

\begin{figure}[p!]
\centering
\includegraphics[scale=0.6]{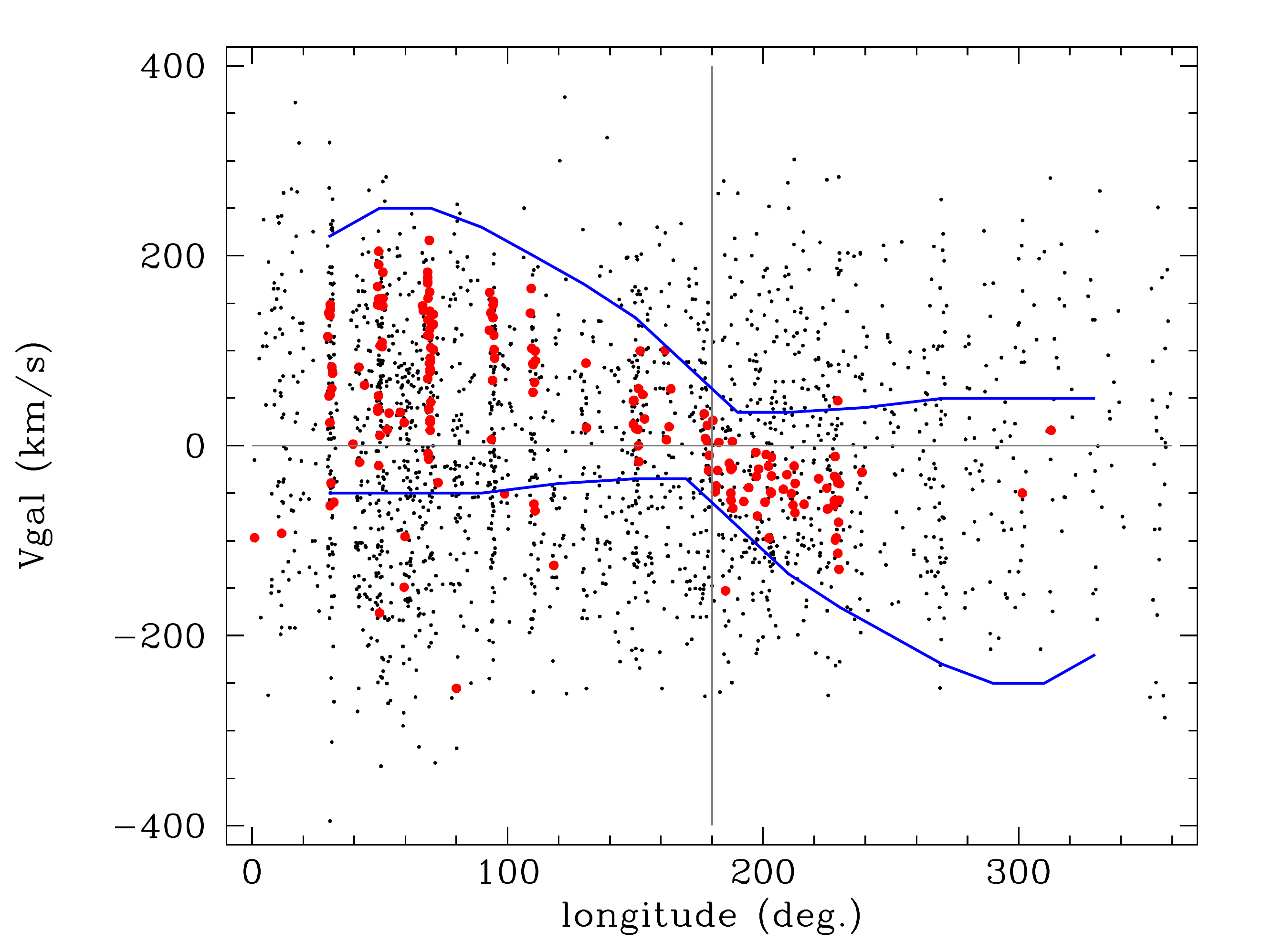}
\caption{Galactic longitude vs radial velocity, corrected for
the motion of the LSR projected on the line-of-sight for all stars
with $R<20$ kpc and $|z|<10$ kpc. Stars with \fe $>-0.8$ are shown in
red. The blue line outlines the region we have chosen to contain
likely disk stars.}
\label{showcuts}
\end{figure}

\begin{figure}[t!]
\begin{center}
\includegraphics[scale=0.6]{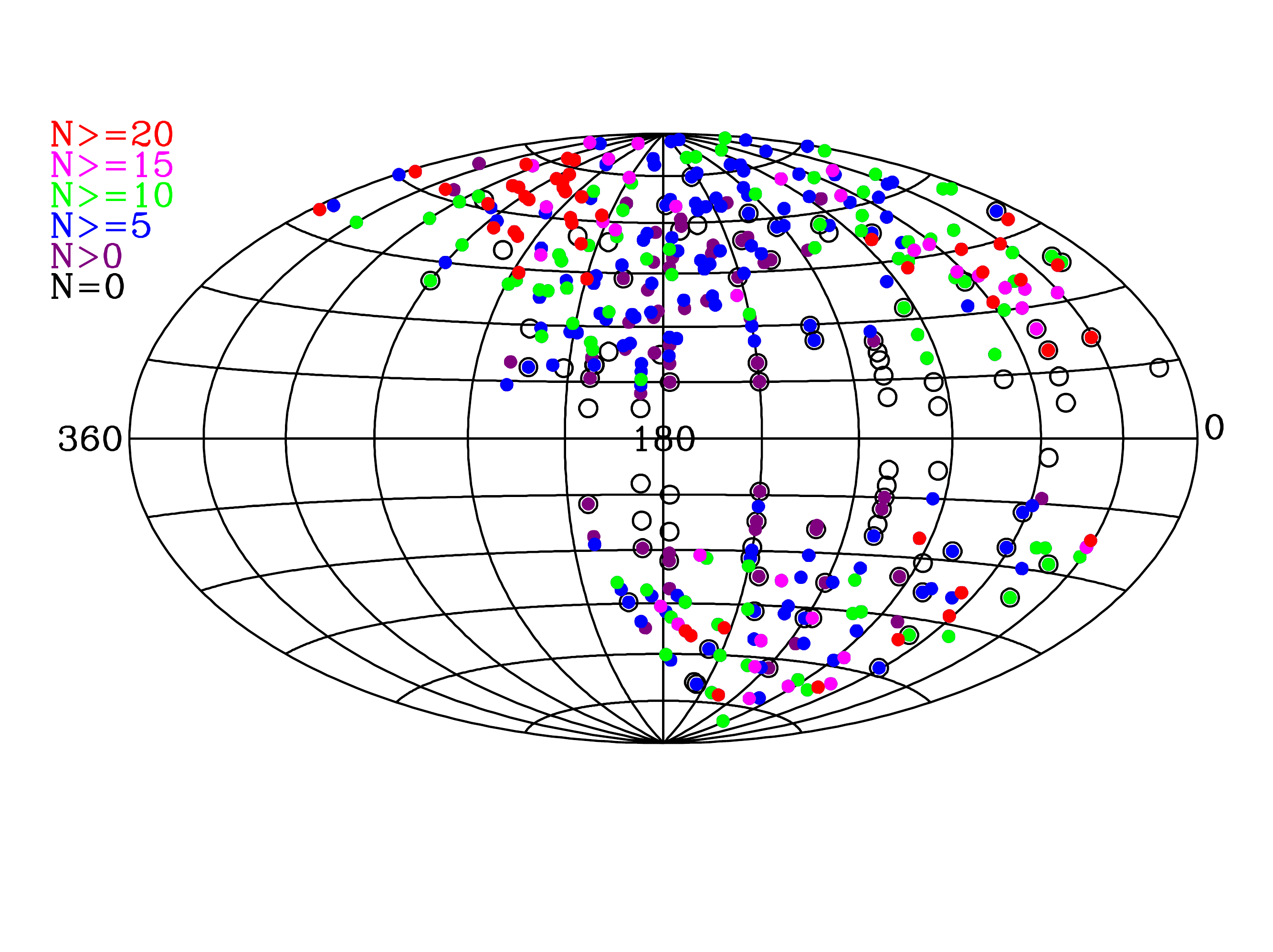}
\end{center}
\caption{The SEGUE plate footprint in the Galactic coordinate system. Each point represents a plate center. The color represents the number of spectroscopically confirmed K giants in our sample on each plate (between 0 and 25), illustrating that some regions on the sky are more dense in K giants than others.   \label{footprint}}
\end{figure}

Our last criterion employs a longitude-velocity plot (often used in
studies of gas dynamics because a distance estimate is not readily
available) in order to require that the star has kinematics consistent
with the disk/thick disk.  Figure~\ref{nodistlv} both highlights the
kinematics of our giant sample to show how the disk/thick disk appears
in these plots, and also shows how bright streams such as those from
the disruption of the Sgr dwarf galaxy will appear. The Figure shows
the radial velocity (corrected to account for the LSR velocity around
the Galactic center) versus Galactic longitude for our enlarged K
giant sample ($\sim$15,000 stars). Here we have included another 10,046
stars not presented in the X14 sample because they are bluer than the
intersection of the horizontal and giant branches and so have two
possible distances. We know that these stars are giants, but do not
know their distances.  Because we do not need the distance of
a star to place it on the longitude-velocity plot, we can increase our
sample size by a factor of three here, and see the kinematic patterns more
clearly. The top panel shows the entire sample. There are two patterns
of clumping in the longitude-velocity plane: one which is most visible
for $l<180$, showing positive velocities which peak near 200 \kms, and
the other which shows negative velocities and is visible for almost
the entire range of longitude.

The second panel shows the $l-v$ distribution of more metal-rich stars
(\fe$>-0.5$) at low latitude ($|b|<35^\circ$).  One would expect
these criteria to isolate stars from the Galaxy's disk and thick disk.
Since Galactic rotation is defined to be positive in the direction
$l=90^\circ$ \citep{binneymerrifield}, we see that the chosen stars have the
correct sign of velocity for disk objects.  The range of $v_{gsr}$
seen for disk stars is a function of both their position along the
line-of-sight and of their population's velocity dispersion. Since
$V_\phi$ is the only component of the velocity of a disk/thick disk
star with non-zero mean, a star whose position places the direction of
$V_\phi$ close to the line-of-sight will show a significant velocity
in the direction of rotation of the Galaxy's disk. Conversely, stars
at the anticenter will have velocity close to zero because $V_\phi$ is
orthogonal to the line-of-sight there, as can be seen in this panel.
As a star's latitude increases, the component of $V_\phi$ along the
line of sight will also decrease because of projection. The velocity
dispersion of the population along the line-of-sight will then
`jiggle' the star's position away from its geometric expectation.

The third panel shows stars from the same latitude range
($b<35^\circ$), chosen to have [Fe/H] less than --1.0. Here we see a
completely different velocity distribution, which is basically
symmetric around zero velocity, except for some substructure likely to
be stellar streams. This symmetry in velocity suggests that the stars
belong to the Galactic halo, which has a near-zero mean velocity
\citep[eg][]{zinn85,fermani12}. We also see little contribution from
the metal-weak thick disk \citep{jen85,mff90,daniela10} in our sample.

Lastly, the bottom panel shows only giants with distances available in
X14, which are chosen to have $R_{gc}>20$ kpc. Since the edge
of the Galaxy's disk is at approximately this distance, we would
expect this criterion to exclude most disk stars and all stars from
the inner halo. The feature which we can see stretching across values
of longitude from 60 to 360 is caused by the various wraps of the Sgr
dSph, and will be discussed in more detail in Section \ref{fof}. Since
the pericenter of the orbit of the Sgr dwarf is at $R_{gc} \sim$ 17
kpc \citep{lm10} and the current position, close to pericenter, is in
the Southern Hemisphere, and thus not accessible to SEGUE, we should
see almost all Sgr members in our sample in this panel.

Figure~\ref{showcuts} illustrates the velocity cut we apply to remove
disk stars from our sample. We show all stars with $R<20 $ kpc and
$|z|<10$ kpc, highlighting those with \fe~ greater than --0.8. It can
be seen that almost all of the red symbols are inside the blue lines
which indicate the velocity cut. 

In summary, we exclude stars as possible disk or thick disk members if they satisfy two spatial criteria (in R and z) {\it and} a kinematical criterion. When these cuts are applied, we end up with 4568 stars in our halo sample.
\FloatBarrier

Spatial substructure is visible in the number of giants found per field. While the actual giant density needs corrections for our observing strategy in a way similar to that of \citet{bovy12} or \citet{katie}, Figure~\ref{footprint} demonstrates that even the uncorrected numbers vary from field to field and between hemispheres in a manner suggestive of spatial overdensities. We note, however, that the fields with low latitude ($|b|<20^\circ$) shown as having no giants were excluded from our giant sample because of their high reddening or because of our disk star exclusion algorithm.

We adopt the SSPP [Fe/H] measurement without applying any further corrections, as H. Morrison et al. (in preparation) have shown that these [Fe/H] values closely correspond with literature values for our calibration sample of star clusters and stars with high-resolution spectroscopy. Typical errors in [Fe/H] are 0.15 to 0.20 dex.

\begin{figure}[t!]
\begin{center}
\includegraphics[scale=0.6]{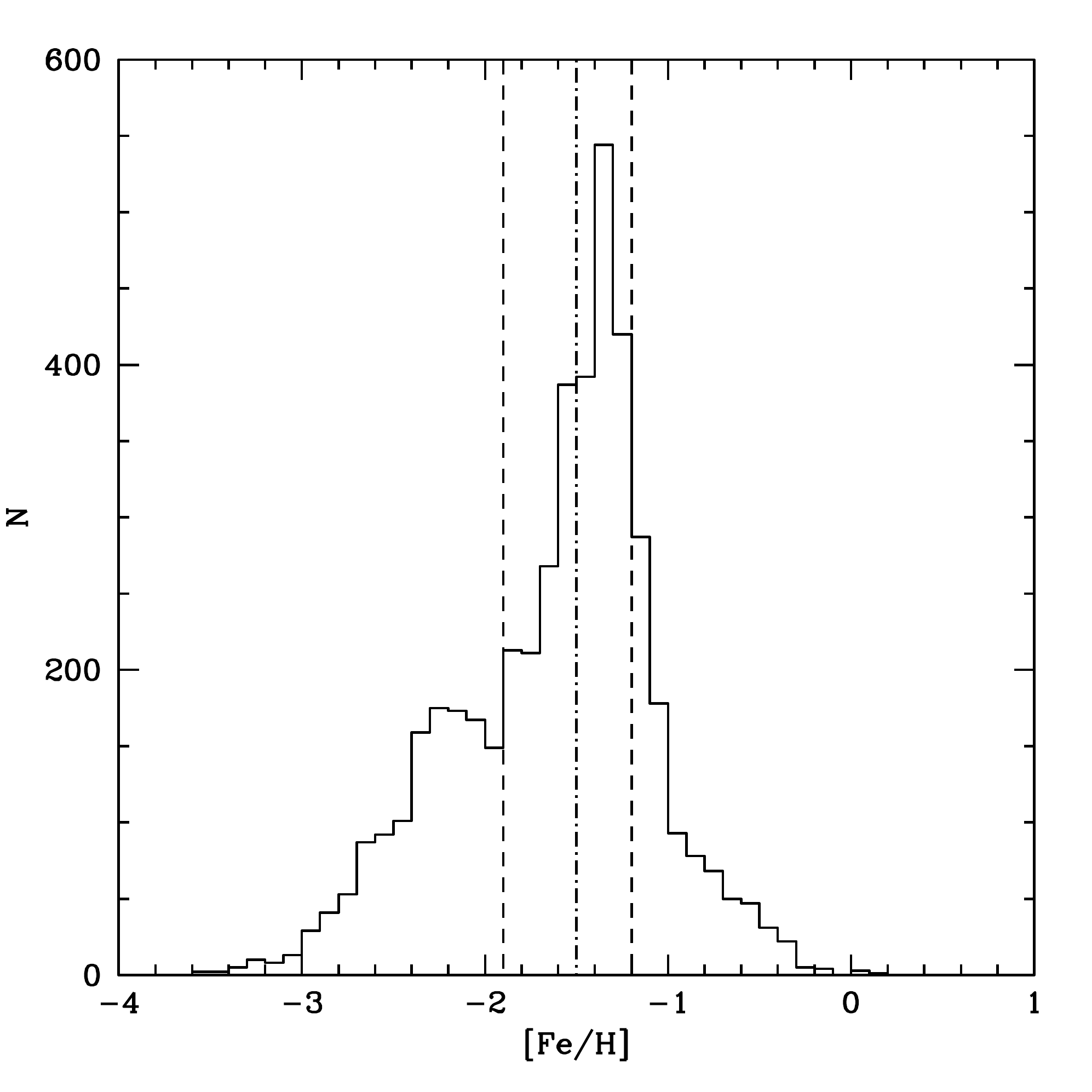}
\end{center}
\caption{[Fe/H] distribution for the spectroscopically confirmed K giant sample with 4568 giants. The vertical dashed and dot-dash lines show the ranges of [Fe/H] we use to create metallicity subsamples below. The dot-dash line represents the median value ([Fe/H] $\approx -1.5$). \label{fehdist}}
\end{figure}

The metallicity distribution of our entire sample of 4568 giants is shown in Figure~\ref{fehdist}. However, we caution the reader against drawing any conclusions about the metallicity histogram of the halo from this diagram, as each target type produces a different metallicity distribution (with the red K giant distribution having a higher mean metallicity, for example), which are not corrected for here. The metallicity distribution of the halo will be the subject of a future paper (Z. Ma et al., in preparation). However, we note that there are a significant number of
stars in our halo sample with [Fe/H] greater than --1.0. Although
traditionally halo stars have been thought to be metal-poor, local
halo samples from earlier work such as that discussed in
\citet{carney96} have so much contamination from the thick disk that
the actual number of halo stars with metallicities greater than --1.0
has not been well constrained.

\begin{figure}[t!]
\begin{center}
\includegraphics[scale=0.6]{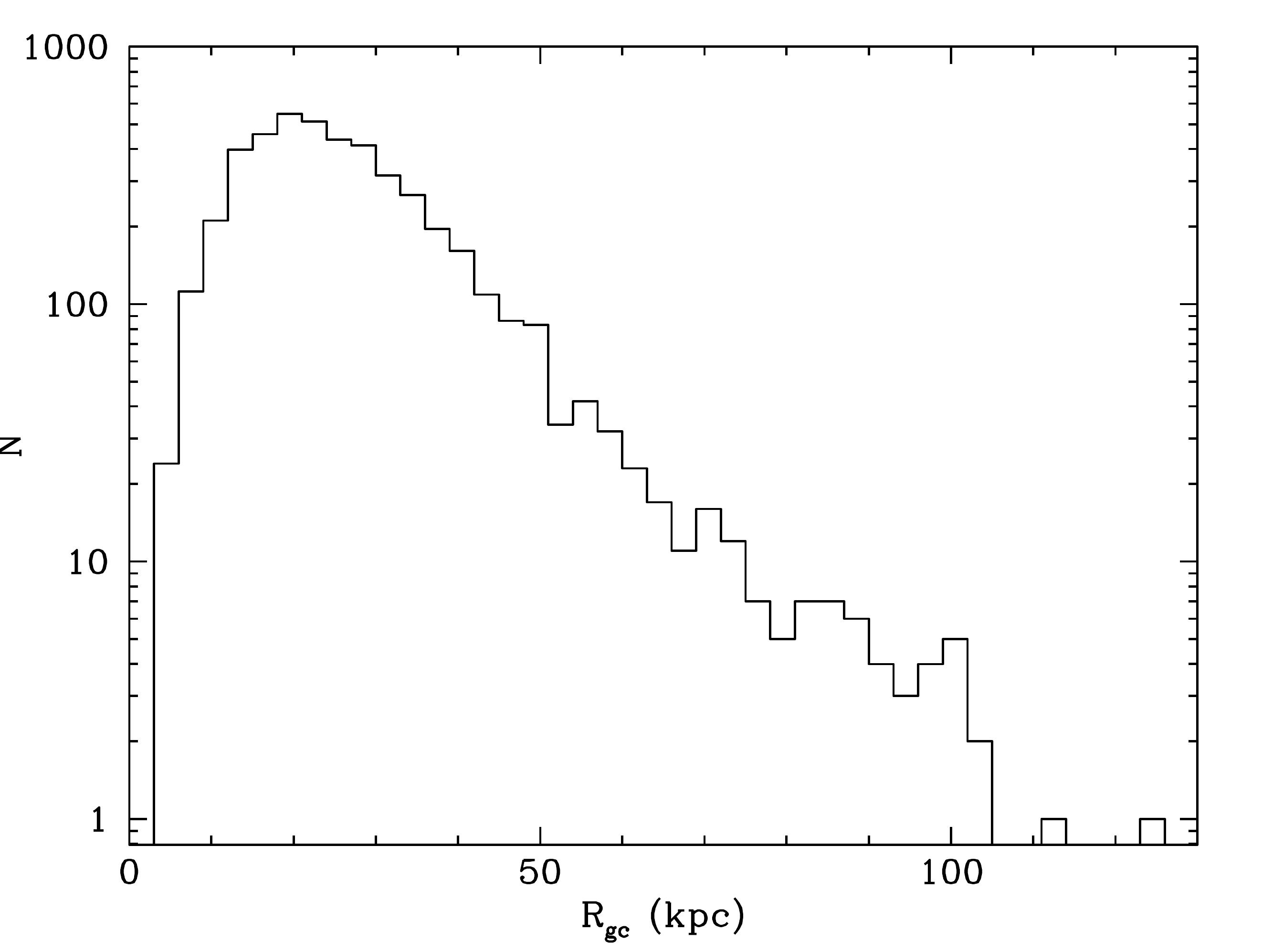}
\end{center}
\caption{Histogram of Galactocentric radius for the 4568 member K giant sample, with distances from \citet{xue14}. \label{kgdist}}
\end{figure}

We used distances for the SEGUE K giants given by \citet{xue14}, calculated using a Bayesian technique, which relies on fiducial giant branch sequences for different metallicities and, importantly, makes corrections for the fact that stars near the tip of the giant branch are much rarer than stars lower down the giant branch. \citet{xue14} show that ignoring this will lead to a bias of 10\% in distance modulus on average, reaching 20\% in some cases. Figure~\ref{kgdist} shows the distribution of distances in our sample: the dataset has significant numbers of K giants at distances of more than 50 kpc, a great improvement on previous halo samples.

\FloatBarrier

\section{The 4distance}\label{4d}
\begin{figure}[t!]
\begin{center}
\includegraphics[scale=0.6]{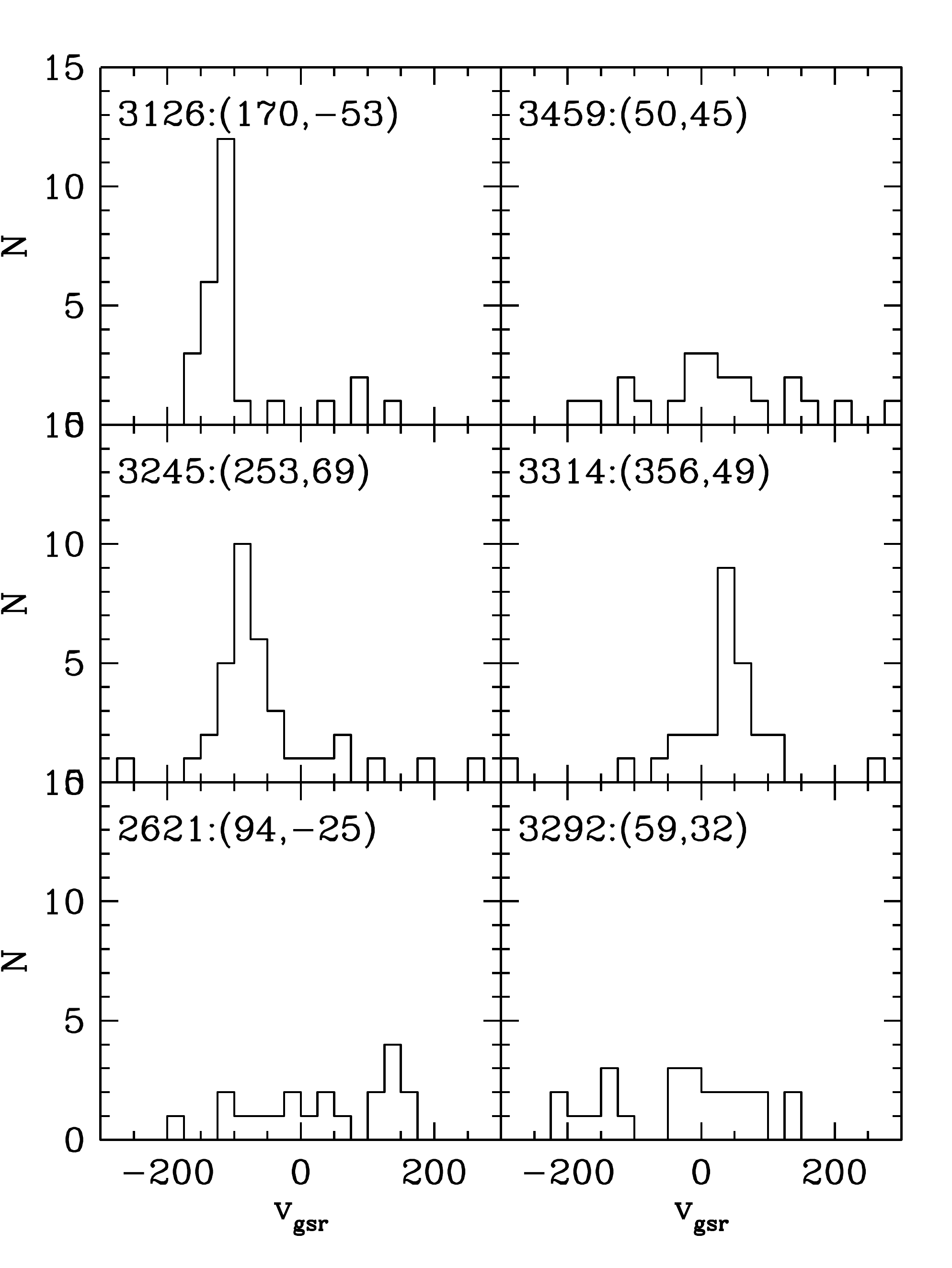}
\end{center}
\caption{Examples of velocity distributions (showing $v_{gsr}$, with the projection of the LSR and solar velocity on the line-of-sight removed) for K giants on individual plates. The label in each quadrant shows the plate number and $(l,b)$ coordinates.  \label{kgvgsr}}
\end{figure}

There have been many approaches to measuring substructure: one of the simplest being simply to count the number of halo turnoff stars. This technique highlights spatial overdensities in the halo, which are particularly apparent in the early stages of a stream's disruption \citep{bullock05}. Our Figure~\ref{footprint} shows suggestions of such spatial overdensities, but would require a proper correction for our complex SEGUE observational strategy to be compared with data such as the `Field of Streams'. Another simple approach is to study the velocity distribution in different fields of a pencil-beam survey, as modeled by \citet{harding01} and implemented by \citet{schlaufman}: deviations from a smooth shape in halo fields are likely to be caused by disrupted streams, and this has the advantage of retaining a signal for later stages of disruption. In Figure~\ref{kgvgsr} we show some typical velocity histograms from giants in SEGUE fields. It can be seen that few of these histograms have smooth or near-Gaussian velocity fields, suggesting a large degree of substructure in our halo sample.

The method that we have chosen to study substructure combines several different dimensions of data: position on the sky, distance, and velocity. As in X11, we use the 4distance method of \citet[][S09 hereafter]{starkenburg}. This metric takes advantage of SEGUE's capability to measure several dimensions in phase space, specifically latitude, longitude, line-of-sight radial velocity, and distance ($l,b,v_{los},d_{sun}$). The 4distance calculates the separation of two stars in each of these four quantities, applies a weighting factor dependent on the measurement range and error, and then produces a dimensionless distance in phase space, 
\begin{equation} \label{pairdef}
\delta_{4d}^{2} = w_{\phi} \phi_{ij}^2 + w_{v}(v_i - v_j)^2 + w_{d}(d_i - d_j)^2,
\end{equation}
\noindent where the angular separation $\phi_{ij}$ is defined by: 
\begin{equation}
\cos \phi_{ij} = \cos b_i \cos b_j \cos(l_i - l_j) + \sin b_i \sin b_j.
\end{equation}

The three weighting factors ($w_\phi, w_d, w_v$) in the 4distance have a two-fold function: to scale the result to physical quantities and to take into account the measurement errors of the data in the metric itself. These weights are defined as
\begin{equation}
w_{\phi} = \frac{1}{\pi^2},  
\end{equation}
\begin{equation}
w_d = \frac{1}{130^2} \frac{(\frac{d_{err}(i)}{d(i)})^2 + (\frac{d_{err}(j)}{d(j)})^2}{2<\frac{d_{err}}{d}>^2},
\end{equation}
and 
\begin{equation}
w_v = \frac{1}{500^2} \frac{v_{err}^2(i) + v_{err}^2(j)}{2<v_{err}>^2}.
\end{equation}	
Since the error in angular position is negligible, it is not included here. However, velocity and distance both often have a significant error component. In our case, distance error scales with distance, but velocity error does not scale with velocity. This is reflected in the weighting factors. The choice of the constant in the weighting factor is determined by the range of the observed quantities and represents the maximum separation between two stars in the 4distance metric. 

\begin{deluxetable}{lr}
\tablewidth{0pt}
\tablecaption{Maximum physical component size for a selected 4distance ($4\delta=0.03$) \label{fdtable}}
\tablehead{Component & Size}
\startdata
$\phi$ $(l,b)$ & $5.4^\circ$ \\
distance (heliocentric) & 6 kpc \\
velocity (Galactocentric standard of rest) & 15 km s$^{-1}$
\enddata
\end{deluxetable}

Table \ref{fdtable} shows some examples of physical sizes corresponding to a 4distance of 0.03. For example, if the 4distance was
calculated for two stars with identical values of l, b and distance, then a difference of 15 km s$^{-1}$ in velocity would produce a 4distance of 0.03. For stars on the same SEGUE plate,
the range in 4distance can be quite large for plates with large numbers of ($>20$) stars. An
extreme example is plate 3245, which has 36 stars and has a minimum
4distance of 0.006, median 4distance of 0.185, and maximum 4distance
of 1.084. For stars separated by a plate diameter (2.98 deg), the
minimum 4distance (equal velocity and distance) is 0.016.

We define a \emph{pair} of stars as being two stars separated by no more than a certain 4distance. We can, in effect, derive a two-point correlation function in a sample by comparing the number of pairs we observe to the number we would expect from a smooth distribution. The exact value we measure is the ratio of the cumulative number of pairs in the \emph{data} with $\delta_{4d}$ less than a certain value to the number of pairs in a ``smooth" sample (we describe the characteristics of a smooth sample in Section \ref{smoothdescription}), which we will refer to as the 4distance ratio. As $\delta_{4d}$ (or the 4distance bin) increases, we expect the 4distance ratio to approach one as the cumulative number of pairs in the data and smooth sample become equal, meaning that at larger separations, there are no additional pairs being added to the cumulative sum. Error bars on our measurements are simply due to Poisson statistics, modeled using a Monte Carlo approach.

%, by taking the standard deviation of a large number of randomly distributed 4distance ratios. These ratios are calculated by adding a random number following a Gaussian distribution of width $\sqrt{N_{pairs}}$ independently to both the number of pairs in the data and number of pairs in the smooth sample.

When small values $(\delta_{4d} <0.10)$ of the 4distance have a relatively large signal, we can infer the existence of streams and other cold substructure in the sample. The 4distance is a robust measurement of the amount of substructure in a population, but is not always an ideal indicator of the presence of streams. The spectrographic plate design of the SDSS forces objects on the same plate to be observed at relatively small angular separations, which causes the bulk of the contribution to the 4distance between two stars on the same plate to be from velocity and distance. The number of pairs will also vary as a function of distance. Whether through a magnitude limit or the halo density distribution function, the number density of stars decreases with increasing distance. This relationship causes a distant star to be less likely to be paired with another, but emphasizes the significance of finding pairs at large distances. 

\begin{figure}[t!]
\begin{center} %this should be a plot of our k giants vs. s09 k giants showing that we get the same answer
\includegraphics[scale=0.6]{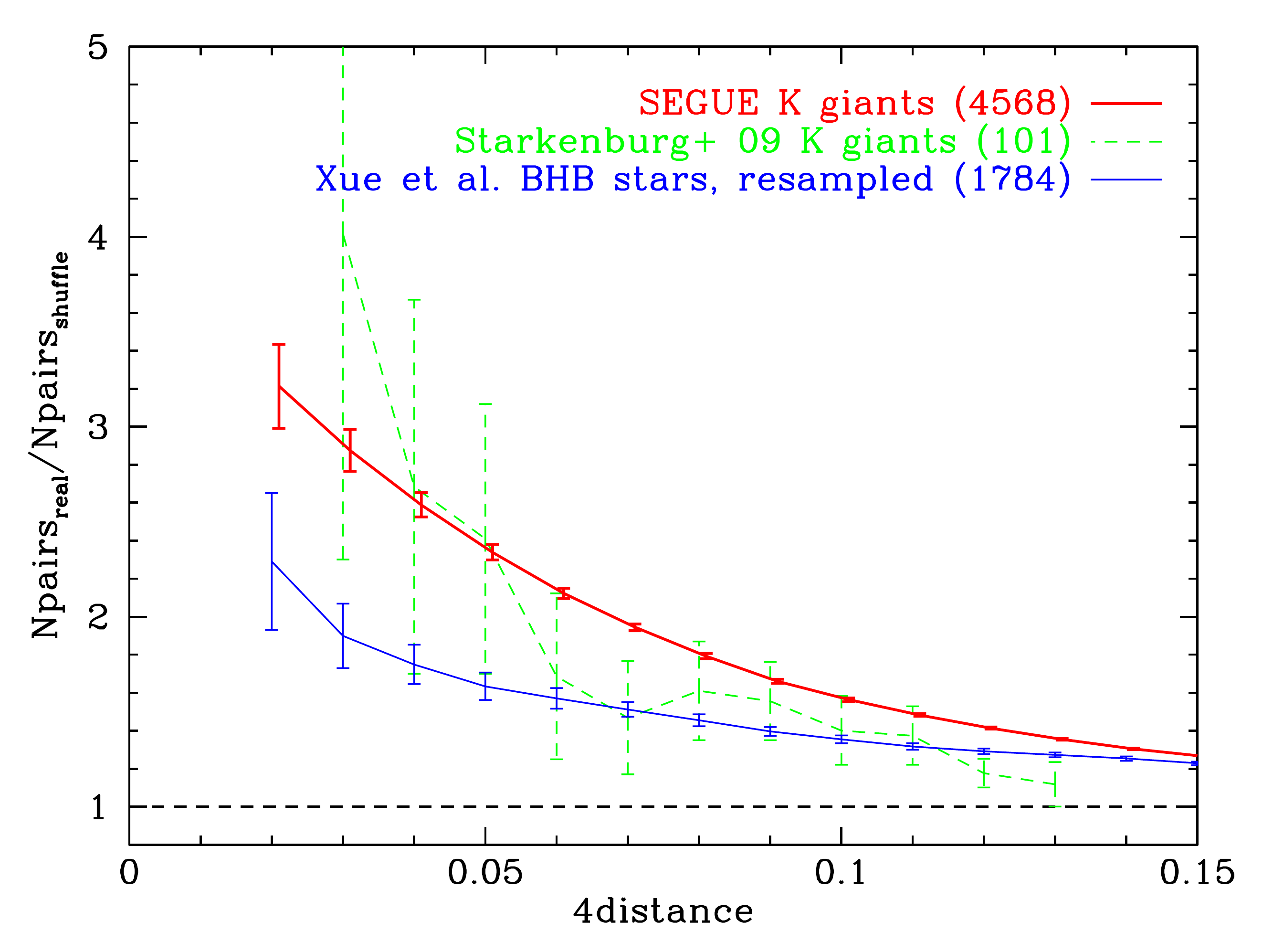} 
\end{center}\caption{4distance measurements for our K giants (red), those of \citet{starkenburg} (green dashed), and a \emph{resampled} set of BHB stars from X11 (blue) using the shuffle method of normalization used by \cite{starkenburg} \label{usvelse}}
\end{figure}

\subsection{Normalization with a Smooth Halo Model} \label{smoothdescription}
\subsubsection{Motivation}
In order to normalize the global indicator of substructure (the $4\delta$ ratio) we need to find the number of 4distance pairs we would \textit{expect} to see in a smooth halo. 
%This is similar to what cosmologists do with the two-point correlation function, when they build a random (uniform) dataset of galaxies to test against (for example, see \citet{zehavi02}). 
\citetalias{starkenburg}, X11, and \citet{cooper11} solved the problem of normalization by randomly shuffling their data. By holding $(l,b)$ constant and independently shuffling $v_{gsr}$ and $d_{sun}$, it was possible to approximate a smooth halo while respecting a survey footprint. We compared the pair counting results for our K giant sample to a shuffle normalization; the results are shown in Figure~\ref{usvelse}. The measurement for our order of magnitude larger sample is consistent with the S09 measurement. We see that, for a given value of 4distance, the K giants have between 1.5 and 2 times more pairs than the BHB stars that have been resampled to select those only inside the SEGUE footprint: they show signficantly more substructure. However, their radial distribution is different, with the K giants stretching to twice the distance of the BHB stars because of their larger luminosity. X11 have shown that the amount of substructure in their BHB sample increases with distance from the galactic center, and we will show that the K giants have a similar property in Section \ref{pairsdistance}. We will further investigate the K giant/BHB differences in Section \ref{kgbhbdiff}.

\begin{figure}[t!]
\begin{center} %this should be a plot of our k giants vs. s09 k giants showing that we get the same answer
\includegraphics[scale=0.6]{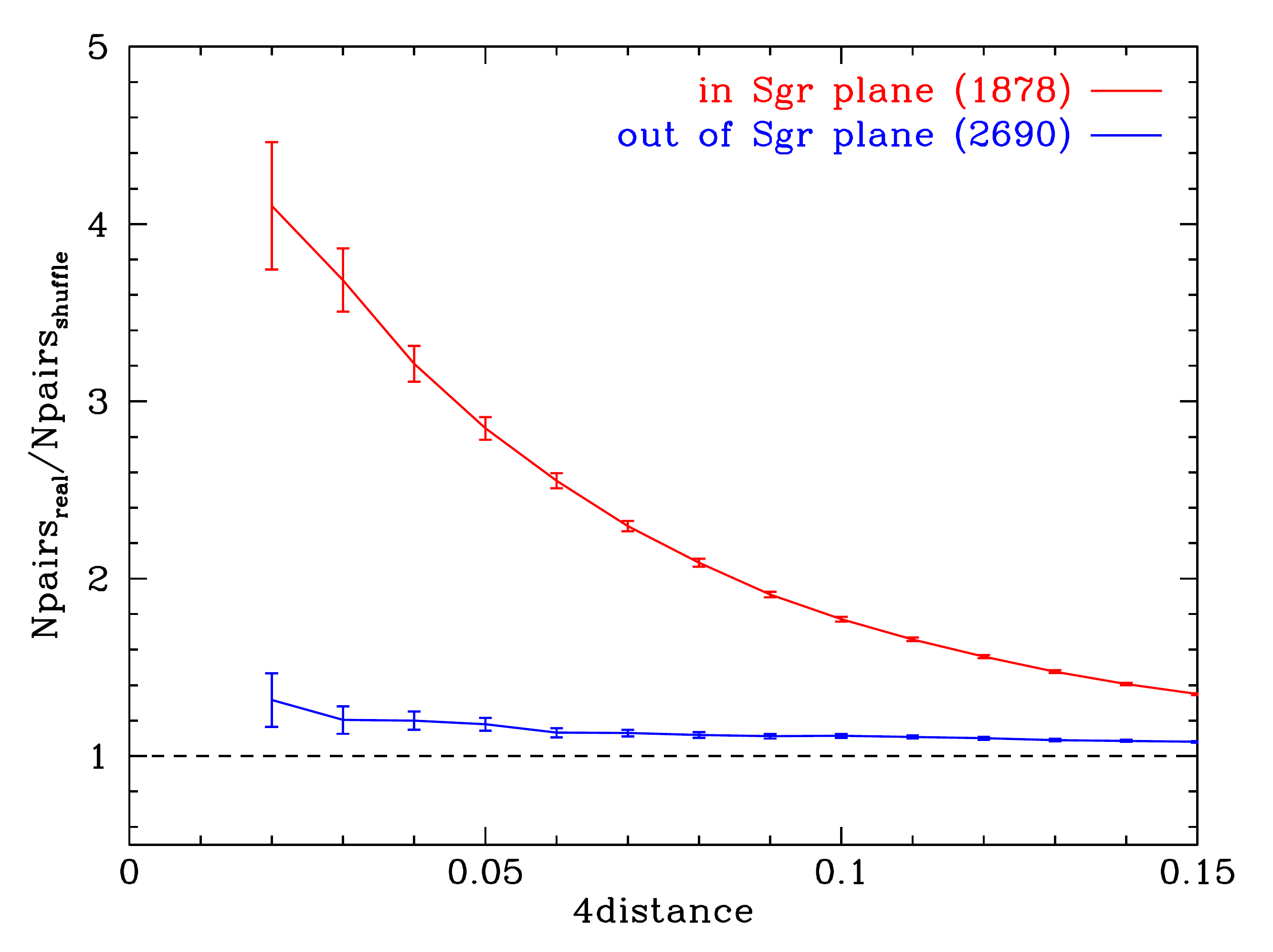} 
\end{center}\caption{4distance measurements for K giants in the Sgr plane ($|B| < 12$; magenta) and out of the Sgr plane ($B > |12|$; cyan) using the shuffle method of normalization used by \cite{starkenburg}. As expected, the stars in the Sgr plane show a much higher substructure signal than stars out of the Sgr plane. \label{pairssgrplane}}
\end{figure}

To show one way the 4distance metric can be used to quantify substructure in the halo, we have divided the K giant sample based on its position in the Sgr coordinate system ($\Lambda$, $B$) described by \citet{majewski03}. Stars with $|B| < 12$ are classified as being in the Sgr plane, while all other stars are outside of the plane. We show the results of 4distance pair counting for these two subsamples in Figure \ref{pairssgrplane}. We obtain the expected result, which is that stars in the Sgr plane show a significantly larger amount of substructure than stars out of the Sgr plane.

However, we note that shuffling normalization method does not account for all aspects of spatial substructure since it leaves the number of stars at each ($l,b$) constant. We know that spatial substructure is quite obvious in large samples (see \citet{belokurov}), so we wish to include spatial position in the metric. SEGUE's pencil-beam spectroscopic footprint has an uneven spatial sampling, so to understand the effect of spatial distribution on the measurement of the 4distance ratio we must ``observe'' a smooth halo model. Further, spatially concentrated structures are more likely to be identified with the 4distance metric, but will be unaccounted for in the pair ratio when using a shuffle normalization. Thus we have chosen to use a smooth halo model for our normalization, instead of following S09 and shuffling velocity and distance in the survey footprint. We will see below that adding the additional spatial information by using a smooth halo for normalization increases the substructure signal by a factor of $\sim$ 1.4 for our sample.

\subsubsection{Assumptions}
Our choice of a smooth halo model for normalization, while adding sensitivity to substructure, requires assumptions about the properties of a smooth halo, the first of which is the density distribution of stars. We use a power-law density distribution, $\rho \propto r^{-3.5}$ \citep[see][]{zinn85,preston91,vivaszinn06}. It is important to use a reasonable estimate for stellar density because the number of observed pairs depends on the density itself (more dense regions will have more pairs), and we do not want to over-count pairs in the smooth halo, as this would decrease the overall substructure signal. We will use a $\rho \propto r^{-3.0}$ halo model in Section \ref{assumptiontests} to test this effect. Next, we assume a functional form for the velocity distribution of the smooth halo. This choice is less critical than the choice of the density distribution because the velocity component of the 4distance is the one with the smallest bin size compared to the overall range. We choose the velocity ellipsoid from \citet{mff90}, $(\sigma_x, \sigma_y, \sigma_z) = (133, 98, 94)$ km s$^{-1}$ in an attempt to realistically model the halo, but make comparisons to a uniform velocity ellipsoid in Section \ref{assumptiontests}.

Our next set of assumptions concern the observational constraints from our survey. The first two relate to the distance distribution of our sample, but the most important of these is the luminosity function. K giants in SEGUE are a subset of target types, and different target types have different distributions in absolute magnitude. This matters because it affects the K giant detection limits in the halo: some target types are redder stars, which can be intrinsically more luminous. Therefore we create an absolute magnitude distribution for the stars to which we attempt to match the absolute magnitude distribution of the model points. We also need to take into account the apparent magnitude and $R_{gc}$ distributions. The reason for this is to be sure that our model points have a similar, but not explicitly identical, distribution to our data. The apparent magnitude distribution is also not flat, so we treat that the same way as the luminosity function, but also ensure that the $R_{gc}$ distribution is similar to that of the data. The simplest of the observational constraints is the SEGUE plate footprint. Here, we only require a model point to be within 1.49 degrees of a SEGUE plate center. Finally, we require the overall number of points to be the same as observed, and the number of each target type the same as observed.

\subsubsection{Technique}
Because the four target categories have different apparent magnitude and luminosity distributions, we generate points for our smooth halo model for each target category separately. Since the smooth halo needs the same number of points as the actual dataset we are testing for substructure, we count the number of stars from each category, and produce a smooth model with the same number of stars. We produce a random sample with density distribution $\rho \propto r^{-3.5}$ within the observed range of $R_{gc}$ for each target category. We then generate an absolute magnitude (uniformly in the range [$1,-3$]), calculate the apparent magnitude directly from the random distance and absolute magnitude, then randomly select model stars based on the actual distributions of apparent and absolute magnitude for that target category, and check that the point falls within the correct $R_{gc}$ bin for that combination of ($l ,b$) position, apparent magnitude, and absolute magnitude. This procedure has the effect of allowing us to generate an $R_{gc}$ distribution similar to that of the data, but drawn from a smooth underlying population. Also, we only keep points which are within the SEGUE footprint on the sky. Each point is then assigned velocities from the \citet{mff90} velocity ellipsoid with $(\sigma_x, \sigma_y, \sigma_z) = (133, 98, 94)$ km s$^{-1}$. We add errors of 20\% in distance and 5 km s$^{-1}$ in velocity to mimic observational errors.

\subsection{Interpretation \& Behavior}
The 4distance ratio is a measure of global substructure in a sample of stars. As it is a cumulative distribution, the ratio will change on a bin-by-bin basis depending on the physical size of the substructure in the sample. If the structures are more concentrated, the ratio curve will be steeper, while a smoother distribution will have a flatter curve. As $4\delta$ goes to one (which is defined by the weighting scheme as the maximum possible separation between two points in one dimension), the ratio should also go to one. A ratio of one (meaning that the number of pairs is equal in the data and smooth halo) indicates that the data is indistinguishable from a smooth halo. At the maximum separation, all pairs in both the data and the smooth sample (which are the same size) should be in the pair ratio. It is possible for the ratio to drop below 1; this is due to the smooth sample being more structured than the data \emph{at that scale}, which is to be expected given the statistical variation of randomly distributed points.

\begin{figure}[t!]
\begin{center}
\includegraphics[scale=0.6]{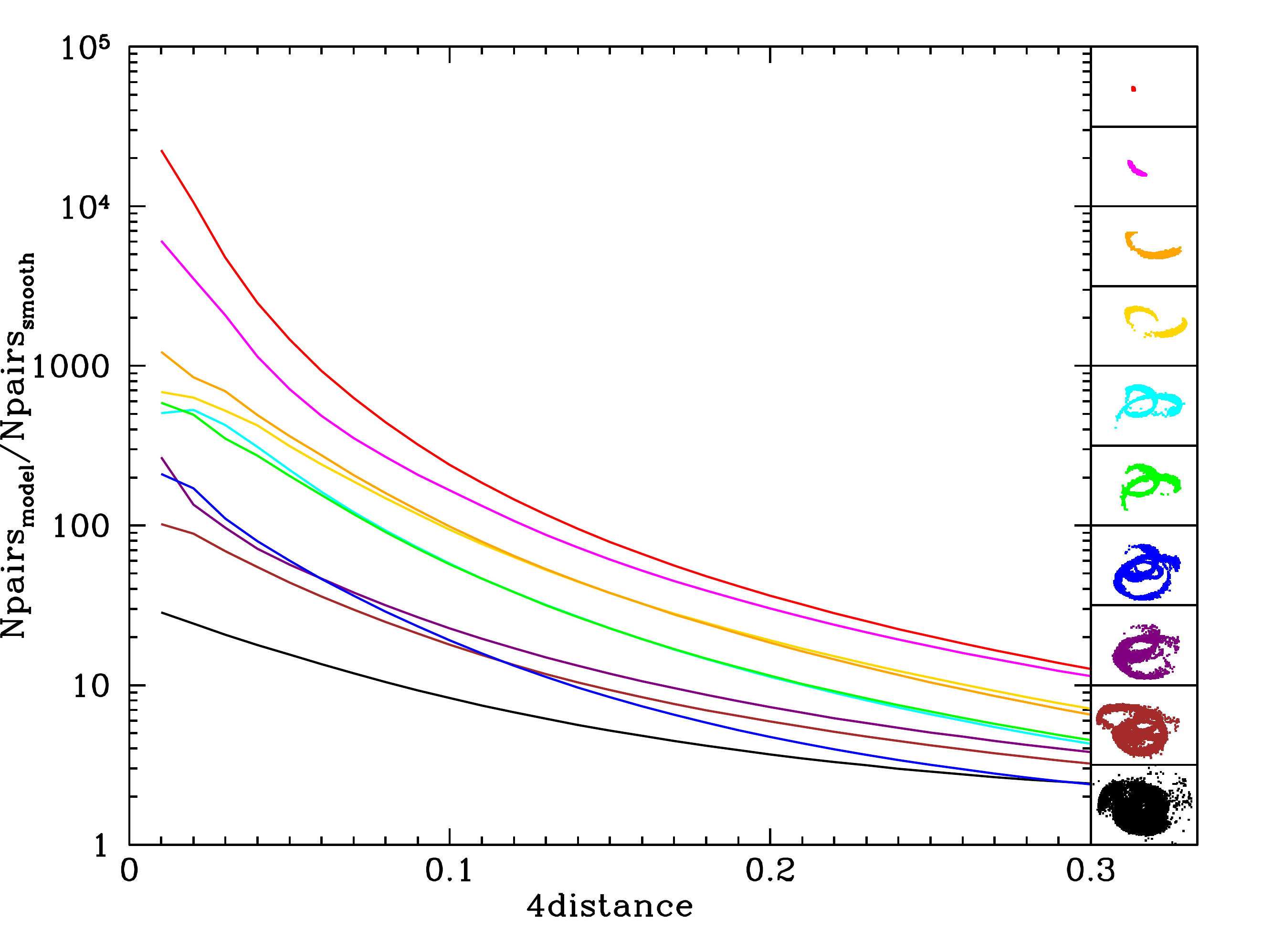} 
\end{center}\caption{4distance measurements for the \citet{lm10} (LM10) model, where debris lost at each pericentric passage is shown in different colors. The ratio at each 4distance bin is calculated by dividing the number of pairs (determined by the 4distance metric, see Equation \ref{pairdef}) in the data by the number of pairs in a smooth halo model (see Section \ref{smoothdescription} for details). The $x-z$ plot for each of LM10's nine wraps are shown in the right hand panels. Starting at the top, these are: currently bound stars, stars lost on the most recent pericentric passage, stars lost on the next most recent passage, and so on. \label{model}}
\end{figure}

Figure~\ref{model} shows the amount of substructure in a model of Sgr disruption created by LM10 when comparing to a $\rho \propto r^{-3.5}$ smooth halo. These models are labeled so that stars lost on each successive pericentric passage of the progenitor are assigned to a different `wrap'. By measuring each wrap separately with the 4distance ratio, we can determine what different kinds of substructure look like in the metric, since the wraps very in their morphology. Figure~\ref{model} clearly shows the difference between the progenitor (in red) and the stars that have had the longest time to mix in the halo (those lost on the first pericentric passage, black). The very dense core of the progenitor has almost a factor of $10^4$ more pairs and a much steeper ratio curve than the more well-mixed stars from the first pericentric passage. 

\begin{figure}[t!]
\begin{center}
\includegraphics[scale=0.6]{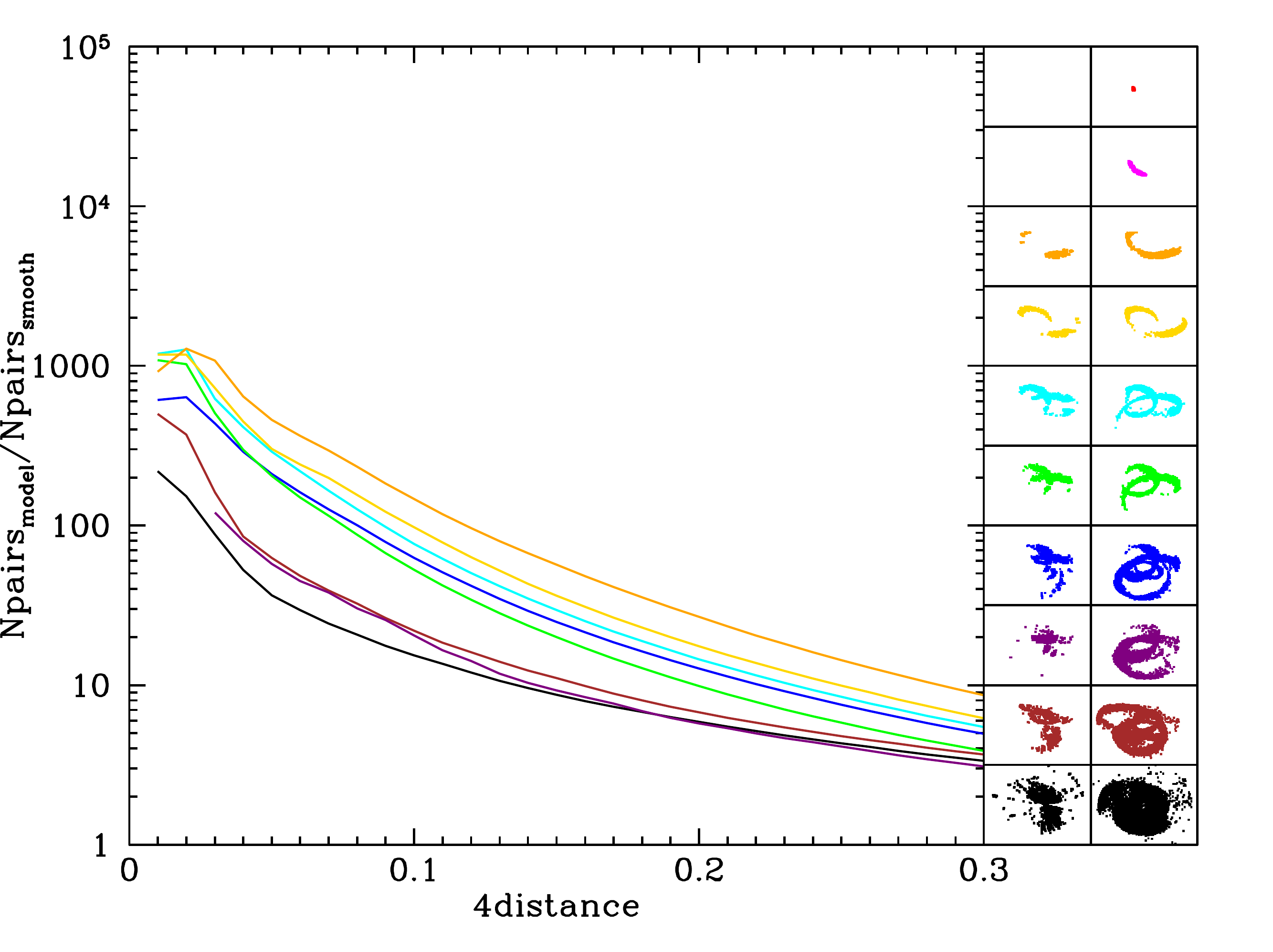}
\end{center}
\caption{4distance measurements for the LM10 model, but observed with the SEGUE footprint. The right hand panel shows the complete model, the middle panel the object in the SEGUE footprint, and the left hand panel the number of pairs compared to a smooth model for the objects in the SEGUE footprint.\label{modelobs}}
\end{figure}

Figure~\ref{modelobs} shows substructure in the LM10 model observed with SEGUE's pencil-beam geometry, again compared to a $\rho \propto r^{-3.5}$ smooth halo. The 4distance measurements follow the same general trend in the observed model as they do without the SEGUE footprint imposed, with only slight differences, indicating that while the survey footprint does add a complication to 4distance measurements, it does not do so in a drastic way. For this reason, however, we caution that we can only directly compare substructure measurements between samples observed with the same survey footprint. Bound stars indicate the highest possible level of substructure, while the lowest observable signal arises from the first wrap, which has had the most time to mix.

\subsection{Variation of substructure signal with model assumptions}\label{assumptiontests}

\begin{figure}[t!]
\begin{center}
\includegraphics[scale=0.6]{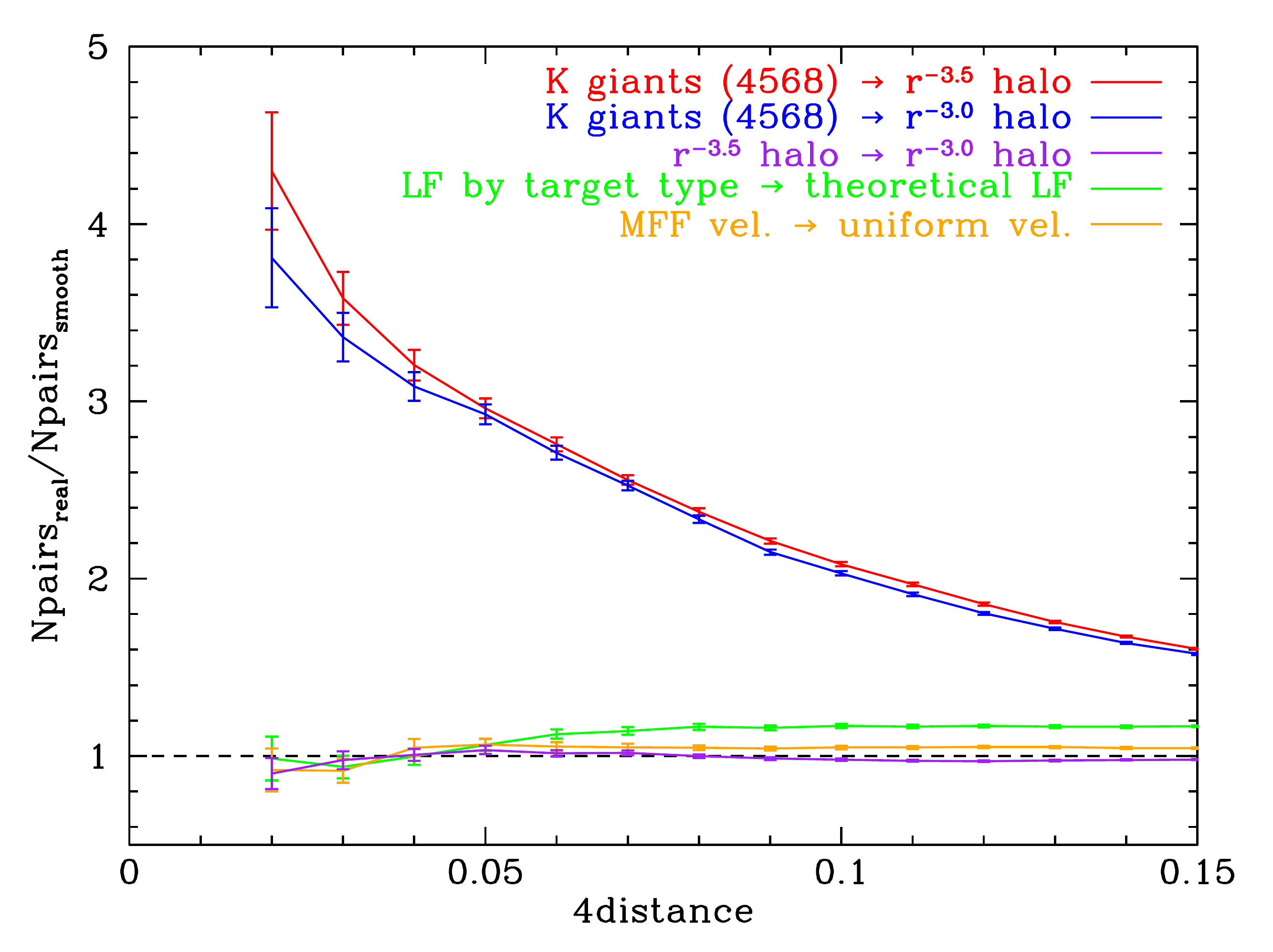} 
\end{center}\caption{Substructure measurements for the full K giant sample, compared to two different model halos ($\rho \propto r^{-3.5}$, red; $\rho \propto r^{-3.0}$, blue). The purple, green, and orange lines show the effect of variations in assumptions on the overall 4distance measurement. The purple line shows a model $\rho \propto r^{-3.5}$ halo compared to a $\rho \propto r^{-3.0}$ halo, the green line shows the  effect of using the observed luminosity function compared to a theoretical luminosity function, and the orange line shows the signal effect of using a \citet{mff90} velocity ellipsoid with $(\sigma_x, \sigma_y, \sigma_z) = (133, 98, 94)$ km s$^{-1}$ compared to a uniform velocity ellipsoid with $(\sigma_x, \sigma_y, \sigma_z) = (100, 100, 100)$ km s$^{-1}$. \label{mateo}}
\end{figure}

Figure~\ref{mateo} shows the result of our 4distance ratio measurement on the full K giant sample, against a number of smooth halo variations. We have measured a stronger substructure signal in K giants using our method with a smooth halo than using the same sample with \citetalias{starkenburg}'s shuffle method, because the smooth halo method allows for spatial clustering to contribute to the signal. 

Tests of the various assumptions in our smooth halo model are depicted in Figure~\ref{mateo}. The purple line shows the effect discussed above, where the centrally denser $\rho \propto r^{-3.5}$ halo has a amount of substructure not found in the $\rho \propto r^{-3.0}$ halo, which is roughly constant over the range of 4distance. Again, this effect is simply due to the fact that increased density leads to more measured pairs. The green line shows the effect caused by the choice of luminosity function. We have chosen to model the luminosity function (LF) by sampling the observed LF for each of our K giant target types, as opposed to using a theoretical LF for the relevant color range for each target type. This choice ensures that our smooth halo model has the same observational properties as the K giant sample. The green line shows that at small 4distance, there is no distinguishable difference between the two, but at larger 4distance, the observational LF counts more pairs than a theoretical LF. Finally, we show that the velocity ellipsoid, shown in orange, has a small effect as well. The cause of this effect is unclear, but could be due to the slightly narrower distributions in $\sigma_y$ and $\sigma_z$ in the \citet{mff90} velocity ellipsoid. Fortunately the change in the 4distance ratio by each of these effects is much smaller than the total signal level in our real sample, as shown by the red and blue lines. We therefore disregard the effect of these assumptions in our final measurements.

Measuring a relatively strong signal using two different normalization methods confirms the presence of substructure in the K giant sample. Furthermore, a strong signal is produced when measured against two separate smooth halo models ($\rho \propto r^{-3.5}$, $\rho \propto r^{-3.0}$). As expected, the $\rho \propto r^{-3.5}$ measurement has a slightly weaker signal than the $\rho \propto r^{-3.0}$ measurement, due to the higher central density in the $\rho \propto r^{-3.5}$ halo, which increases the number of smooth pairs, driving down the overall measurement. 

\subsection{Friends-of-Friends} \label{fof}
Determining a measurement of overall substructure in the halo is useful, but for certain purposes we would like to identify specific structures. Following S09, we use the 4distance in combination with a friends-of-friends (FoF) algorithm to obtain a local measure of substructure by linking stars into groups. These groups are stars with similar characteristics. In a metric system, two stars are associated if they are within a certain linking length, the maximum separation that two stars can have to be identified associated in a given metric. Recalling that a 4distance size contains information about sky position, velocity, and distance, it is useful to review what that means physically. Our chosen linking length of $4\delta = 0.03$ corresponds to physical sizes in each dimension that can be found in Table \ref{fdtable}.

FoF works by drawing a circle around each point, with a radius of one linking length. If the circle contains any other points, those points are in the group. Once a point is included in the group, a circle is drawn around it, and points inside that circle can now be added. The group is complete when no more points can be added. Since the 4distance involves more than just physical position, our circles need to be drawn in more than just spatial coordinates, but this highlights the power of the 4distance metric--the fact that it uses velocity information to find stars that are moving in the same direction.

Of additional concern when choosing a linking length is the pencil-beam geometry of the SEGUE survey. We want groups to extend across more than one plate pointing, but at the same time, we do not want the FoF algorithm to be too generous in linking spatial positions. In other words, while we desire the ability to connect nearby plate pointings, the deciding membership factor should be distance or velocity. Our choice of linking length of $4\delta = 0.03$ takes these criteria into account. The maximum angular separation of 5.4$^\circ$ allows for stars on nearby plate pointings to be included as possible group members, but will require them to have  similar distances and velocities to be members of the same group. We will see below (Figure~\ref{groups10}) that each Sgr stream wrap is mapped into a number of groups.

\FloatBarrier
\section{Results on Measuring Substructure in K giants} \label{4dresults}
We have confirmed that the Milky Way halo is highly structured in K giants, but wish to make more meaningful conclusions about the nature of the halo. Given our large sample size compared to the previous K giant sample of \citetalias{starkenburg}, we are able to create subsamples that will allow us to explore both the inner and outer halo, as well as ranges in metallicity.

\subsection{Variation with Galactocentric Radius} \label{pairsdistance}
\begin{figure}[t!]
\begin{center}
\includegraphics[scale=0.6]{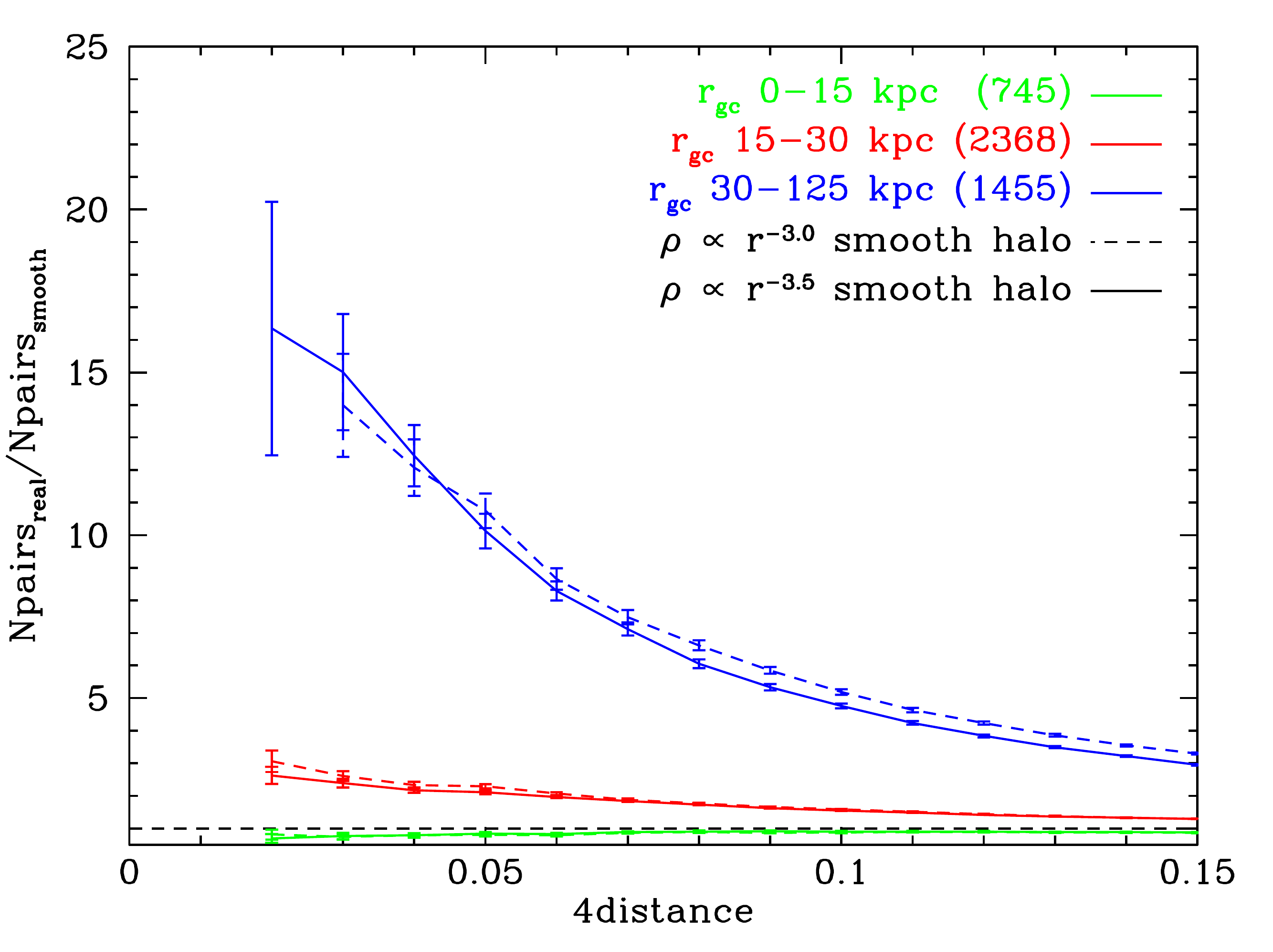}
\end{center}
\caption{4distance measurements for SEGUE K giants, divided into distance ranges in Galactocentric radius. The innermost subsample (shown in green), from $R_{gc}=0-15$ kpc, exhibits the least amount of substructure, while the outermost subsample (shown in blue), $R_{gc} > 30$ kpc, shows the most. Red shows the subsample with $R_{gc} = 15-30$ kpc. We also present two separate smooth halo normalizations in this figure ($R^{-3.0}, R^{-3.5}$) in dashed and solid lines, respectively. \label{kgnrf}}
\end{figure}

We might expect the inner halo to show less substructure than the
outer halo for two reasons. First, any smooth component of the halo
(formed, for example, by violent relaxation in the early stages of
halo formation) would be found in the inner regions. Second, accreted
stars will likely phase-mix faster in the inner halo due to the
shorter dynamical times there. This expectation is largely borne out
in the simulations of \citet{bullock05} and \citet{cooper10}. In both papers the
majority (but not all) of the realizations show more substructure in
the outer than the inner halo \citep{xue11,cooper11}.

In our data we see substructure increasing monotonically with distance 
(see Figure~\ref{kgnrf}), with the most distant stars (with $R_{gc}$ from 30 to 120
kpc) showing the strongest signal, $\sim$5 times more pairs than the
halo stars with $R_{gc}$ between 15 and 30 kpc. While stars at these
intermediate distances still show significant substructure, this is
not the case for stars with $R_{gc}$ less than 15 kpc. However, there
is abundant evidence for substructure in this region from other
studies using either velocities \citep{schlaufman} or the full 6-D
phase space information \citep[e.g.][]{helmi99}. This highlights the
fact that the 4distance is a relatively rough measurement of
substructure.

\subsection{Metallicity Dependence} \label{pairsmetal}
\begin{figure}[t!]
\begin{center}
\includegraphics[scale=0.6]{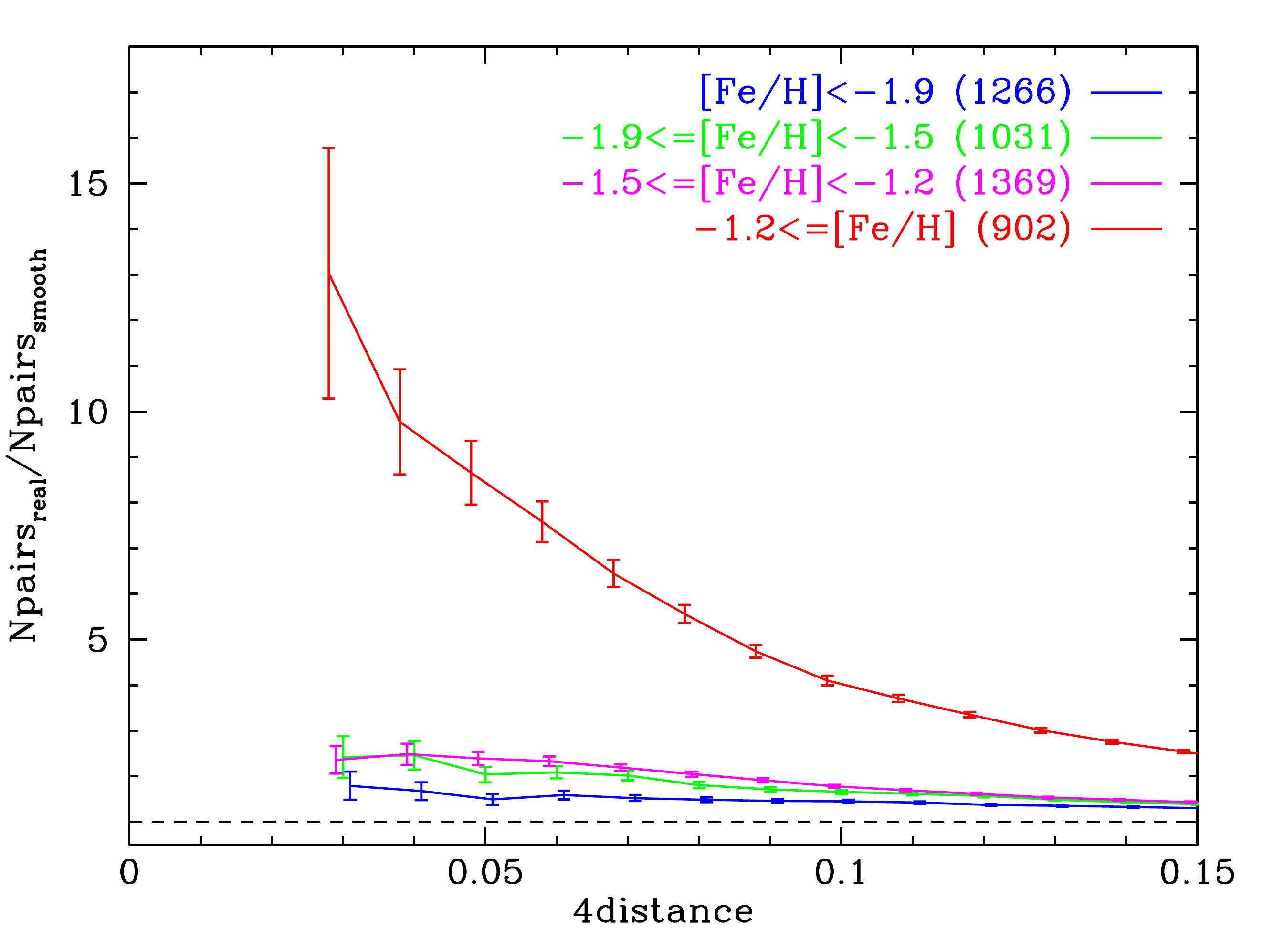}
\end{center}
\caption{4distance measurements for SEGUE K giants, divided into abundance ranges. The median metallicity of our K giant sample is [Fe/H] $\sim -1.5$. The red and blue lines represent the top and bottom 25\% of the metallicity range, respectively. More metal-rich K giants show a high level of substructure, while the two metal-poor categories show similar, but lower, levels.\label{kgfeh}}
\end{figure}

The use of K giants as tracers allows us to check explicitly for a
dependence of substructure on metallicity, since it is considerably
easier to obtain accurate \fe\ values for K giants than for BHB stars
due to the much stronger metal lines in K giants. (Morrison et al, in
preparation, demonstrate this for the SSPP metallicity measures.) This
is of particular interest in the context of the disruption of accreted
satellites into the field halo because of the mass-metallicity
relation \citep{massfeh} which allows us to infer that if a low-mass object is
accreted, its stars are likely to be metal-poor.

Figure~\ref{kgfeh} shows the global substructure measurement for the sample divided into [Fe/H] ranges. The most metal-rich group {\fe $>-1.2$) shows a very strong
substructure signal, with $\sim$ 8 times more pairs than a smooth halo
at 4distance = 0.05. The two groups with intermediate metallicity
(\fe\ between --1.2 and --1.9) have $\sim$ 2.5 times more pairs than
the smooth halo at this value of 4distance, while the most metal-poor
group (\fe$<-1.9$) shows the most subtle signal, with $\sim$ 50\% more
pairs than the smooth halo.

\subsection{Comparison of BHB and K giant Substructure} \label{kgbhbdiff}
We can use a comparison between BHB stars and K giants to investigate the properties of stellar populations of different age, because globular clusters with similar metallicity but blue horizontal branches are 1.5-2 Gyr older than those with red horizontal branches \citep{dotter}. Because stars of all ages traverse the red giant branch, while only older stars of the same metallicity will become blue horizontal branch stars, any differences in substructure between the two samples is likely due to age.

Two photometric surveys have led to claims that BHB stars show less spatial substructure than the overall halo population. The first \citep{bell08} counted numbers of main sequence turnoff stars in SDSS photometry covering distances from 7--35 kpc, and showed that there was significant substructure.  \citep{alys11} used a photometric technique to identify BHB stars in SDSS photometry, covering a distance range from 10–- 45 kpc, but using a different method and over spatial scales different from those in \citet{bell08}. Each photometric method has weaknesses which may lead to them underestimating substructure: the distance measures to turnoff stars are not very accurate (Bell et al. estimate 0.9 mag. scatter) and this distance error will smooth out the ``lumpiness'' of spatial substructure along the line of sight. The BHB star sample suffers from some contamination by foreground blue stragglers of the halo, and since X11 show that the substructure in BHB stars increases with galactocentric distance, this too will have a smoothing effect on the amount of substructure.  Thus, although it is clear that the photometric BHB sample of \citet{alys11} shows less substructure than the \citet{bell08} main sequence sample, it is not clear whether this is due to contamination of the BHB sample by inner halo blue stragglers or to an actual difference in substructure between the two tracers.

There is also some indirect evidence for a difference in substructure signal which is arrived at via a comparison with the models of \citet{bullock05}. \citet{bell08} compared their spatial substructure method with the BJ05 models and found good overall agreement, while X11 compared the BHB substructure signal (in both velocity and position) with the same models and found less substructure than predicted by the models. However, since the two comparisons were made of a different signal (spatial vs spatial plus velocity) and over a different distance range (out to 35 vs 60 kpc), and we have seen that substructure in both BHB stars and K giants varies with distance, we feel that this result is not conclusive.

Clearly the spectroscopic samples of both BHB stars and K giants provide a safer measure of substructure, without the smoothing effects discussed above. Our original calculation of the amount of substructure in the K giant and BHB samples using the ``shuffling'' method of normalization (shown in Figure \ref{usvelse}) showed significantly more substructure in the K giants. However, since substructure in both samples increases with galactocentric radius, a correct comparison will limit the K giant sample to the smaller distance range probed by the BHB stars. Figure \ref{kgbhbnew} shows the result of this comparison: we see that the difference between the two stellar types has largely disappeared, and within the errors we see no significant difference in substructure using this method. 

We thus see no conclusive evidence so far of different substructure signals in K giants and BHB star samples. It would be interesting to see a full analysis of both BHB and K giants using a smooth halo normalization as this would be more sensitive, because it also takes into account the spatial variation of numbers across the sky. In addition, more exploration of the relative contribution of BHB stars and K giants in the Sgr stream would be illuminating, but both are outside the scope of this work.

\begin{figure}[t!]
\begin{center}
\includegraphics[scale=0.6]{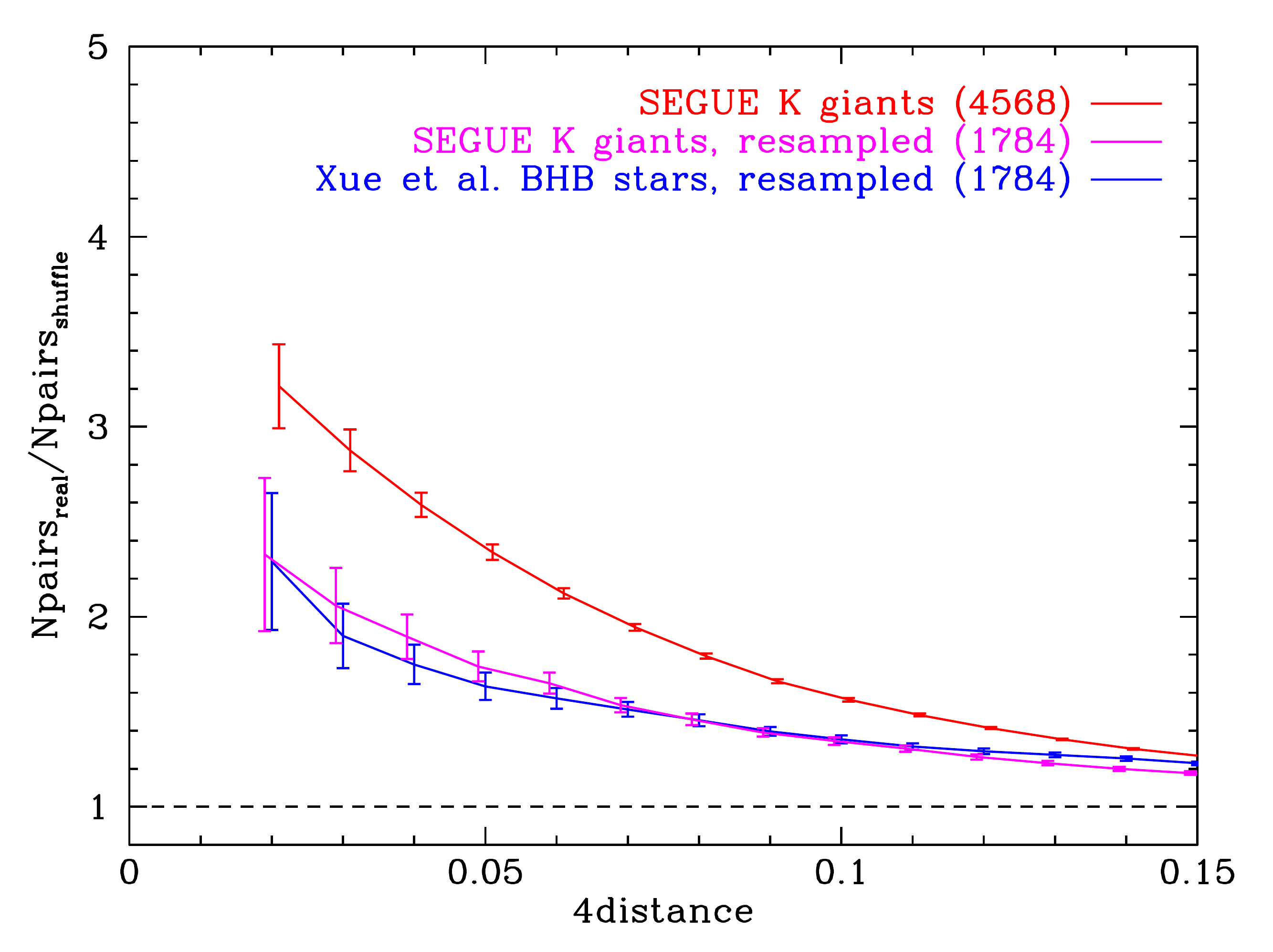}
\end{center}
\caption{4distance measurements for SEGUE K giants (magenta) and X11 BHB stars (blue), resampled to have matching distributions in $R_{gc}$. We also show the 4distance measurement for the intact K giant sample for reference (red). When resampled in this manner, K giants and BHBs show no significant evidence for differing amounts of kinematic substructure.\label{kgbhbnew}}
\end{figure}

\subsection{Distance or Metallicity?}
Since the optical luminosity of a K giant increases as its metallicity
decreases, a survey such as ours with a limiting magnitude will find
only metal poor stars in the most distant regions, and only
relatively metal-rich stars in its most nearby ones. 
We see in Figure~\ref{fehdistsgr} how the survey selection effects play into the
distribution of distance and metallicity in our sample: the most metal
rich stars cover a smaller distance range than the most metal poor ones.

\begin{figure}[t!]
\begin{center}
\includegraphics[scale=0.6]{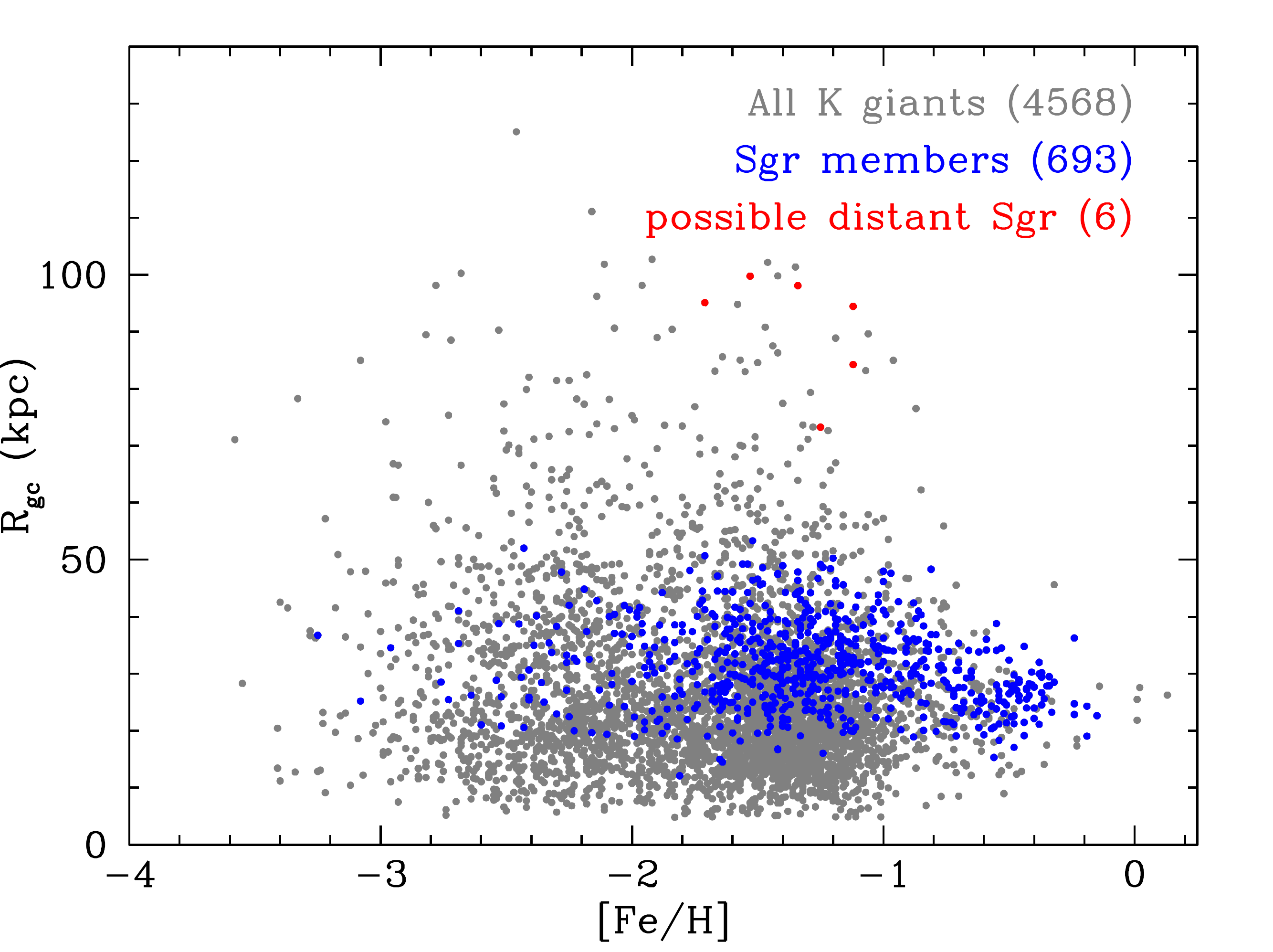}
\end{center}
\caption{[Fe/H] vs distance for our K giant sample. The effects of SEGUE
targeting and magnitude limits can be seen: there are no stars with
\fe $> -0.7 $ more distant than 50 kpc, while the most distant star
has \fe $< -2.4$. Stars which are part of groups classified as
`definitely' belonging to Sgr (see Section 5 below) are shown with blue
points, and stars belonging to a possible distant Sgr group are shown with red points. \label{fehdistsgr}}
\end{figure}

If we saw more substructure in metal-poor and more distant stars we
would need to worry about the above degeneracy, but in fact our results
show the opposite behavior: the most metal-rich and the most distant stars show the most
substructure.

We find that more metal rich stars have more substructure, but what does this imply for halo formation? Though its disruption is ongoing, the Sgr dwarf is a quite massive satellite (a recent estimate gives the current core mass at $\sim 5 \times 10^8$ M$_{sun}$ and intial virial mass at $\sim 1 \times 10^{10}$ M$_{sun}$, placing Sgr among the most massive Local Group dwarf galaxies \citep[see][]{mateo98,lokas10} and has also been observed to be among the most metal-rich of the MW satellites \citep{mateo98}. Given that massive satellites are more metal-rich \citep[see][]{massfeh}, it seems likely that the large amount of metal-rich substructure in the sample can be attributed to Sgr, especially since we have taken pains to remove the disk from the sample and that Sgr is such a visible contributor to spatial substructure in the Field of Streams. It should be noted, however, that the other metallicity ranges do show a significant amount of substructure. We will attempt to quantify the contribution of Sgr and other streams to this substructure below.

\section{FoF results} \label{fofresults}
\citetalias{starkenburg} find eight total groups, and only one group larger than two stars, which they suggest is associated with Sgr. Our results with a larger sample (4950 to their 101), yield a much larger number of groups; our data contain well over 100 groups larger than three members, and more than 20 groups larger than 10 members in size. About $38\%$ of our K giant sample is in a group of four or more members, $\sim 4\%$ is in a group with three members, and $\sim 8\%$ is in a group with two members (a pair). This then leaves $\sim 50\%$ not in a pair or a group. Of S09's 101 stars, $\sim 14\%$ were in a pair, and only $\sim 5\%$ are found in a group of 4 or more stars, clearly illustrating that the power of the 4distance and FoF algorithm increases with sample size. 

This large number of groups allows us to more deeply examine the properties of structures in the Milky Way halo. Significant group detections found via the FoF method can be found in Table \ref{foftable}. Figures \ref{groups10} (groups with ten or more members) and \ref{groups49} (groups with between four and nine members) show these groups in context in four dimensions of phase-space. The largest groups appear to be primarily associated with two large structures, one each in the Northern and Southern Galactic hemispheres. These are, respectively, the Sgr leading and trailing streams, which will be discussed in the next section. The diversity of substructure in the halo becomes more apparent in Figure~\ref{groups49}, which shows smaller groups. Additionally, most of the groups in Figure~\ref{groups49} are located in the northern Galactic hemisphere. 

%[There should be a citation here that we have seen this before, but I can't find it.]

Unfortunately, the FoF technique suffers from the same problem that afflicts the overall pair ratio measurement: there are more groups in regions with ($r_{\odot} \leq \sim 20$ kpc), due to the higher star density coupled with fixed linking length. In this case, this means that the regions that are relatively close to the Sun and Galactic center will appear to have more groups than more remote regions. 

\begin{figure}[t!]
\begin{center}
\includegraphics[scale=0.6]{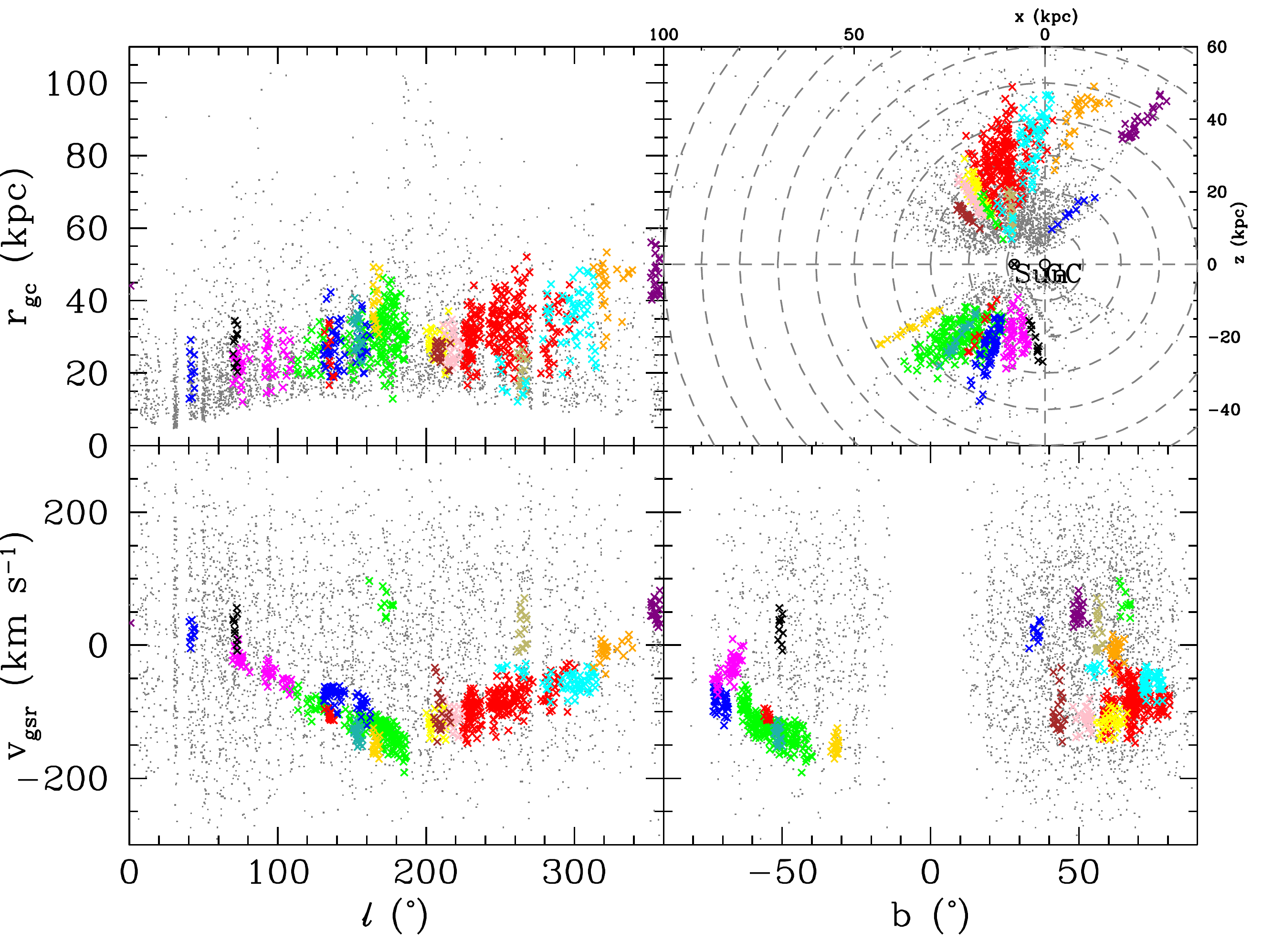}
\end{center}
\caption{The FoF (friends-of-friends) groups greater than or equal to 10 members in size, shown with the line-of-sight velocity (corrected for the solar and LSR motion) $v_{gsr}$, against Galactic latitude $\sc{l}$ (lower left), Galactic latitude $b$ (lower right) and with longitude plotted against Galactocentric radius $R_{gc}$ (upper left). The upper right panel shows the $x-z$ plane, which is close to Sgr's orbital plane. Dashed lines are every 10 kpc in $R_{gc}$. K giants are shown as colored crosses, with each color representing a different group (color repeats between groups should not be construed as an indication of group membership, but are merely caused by a limited number of colors). The full K giant sample is represented by gray dots in the background. Nearly all of these larger groups are associated with Sgr.\label{groups10}}
\end{figure}

\begin{figure}[t!]
\begin{center}
\includegraphics[scale=0.6]{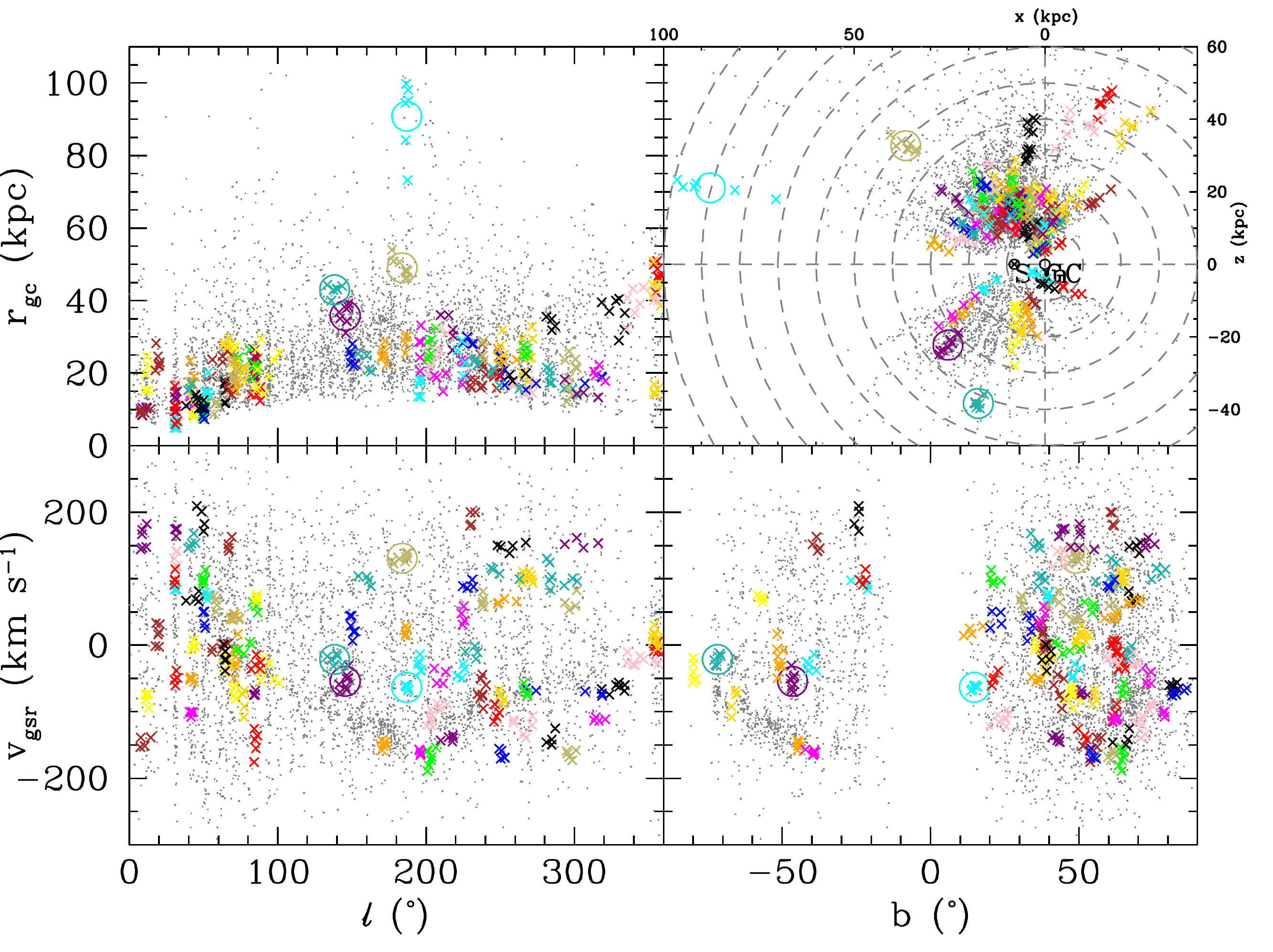}
\end{center}
\caption{The FoF groups between 4 and 9 (inclusive) members in size, shown in 4 different sections of phase space as described in the caption for Figure~\ref{groups10}. K giants are shown in colored crosses, with each color representing a different group (color repeats between groups should not be construed as an indication of group membership, but are merely caused by a limited number of colors).  The full K giant sample is in greyscale in the background. The majority of these small groups fall in the northern Galactic hemisphere. Circled are the four notable non-Sgr substructure groups discussed in Section \ref{coolstuff}: gold, Orphan stream; purple and teal, Cetus Polar Stream; cyan, possible Sgr stream members at 90 kpc. \label{groups49}}
\end{figure}

\subsection{Attributing Groups to Known Substructure: Sgr}\label{sgrboxes}
By comparing to models and observations, we can identify groups associated with known substuctures. The most important of these substructures is the Sgr stream. In fact, the Sgr stream can be seen clearly in a simple longitude-velocity plot even without using the FoF algorithm. Since Sgr streams dominate substructure in the Field of Streams and are the best-studied streams in the halo, we begin with quantifying their contribution to our sample. We need a reproducible technique to identify substructure belonging to Sgr, so we adopt the LM10 model as a way to define regions where Sgr streams should be observable, under the assumption that the LM10 models are an accurate representation of the spatial and kinematic properties of the Sgr streams. By observing the models using the SEGUE footprint (see Figure~\ref{modelplates}), we see the expected appearance of Sgr in a large--scale survey. In addition to LM10's recommendation that only the five most recent pericentric passages are used, distance and velocity errors drawn from Gaussian distributions with sigma values of 20\% and $\sim$ 5 km s$^{-1}$, respectively, have been added to each model point that falls within $1.49^\circ$ of a plate center. By constructing boxes around the observed model, we create the regions  used to identify members of Sgr. We use a simple system to determine membership. There are boxes in four separate dimensions $(l-v_{gsr}, b-v_{gsr}, l-z_{gc}, x_{gc}-z_{gc})$, shown in Figure~\ref{modelboxes}. If 60\% or more of a group's stars fall into the box in all four plots, then the group is assigned to Sgr (referred to hereafter as ``definitely Sgr''). A significant number of groups are identified as Sgr groups, and the majority of them have a large number of members, which means that the most obvious substructures in the halo are associated with Sgr. The width of the potential observed streams in the model is quite large, on the order of 10 kpc or 100 km s$^{-1}$, because Sgr is a massive dSph.

\begin{figure}[t!]
\begin{center}
\includegraphics[scale=0.6]{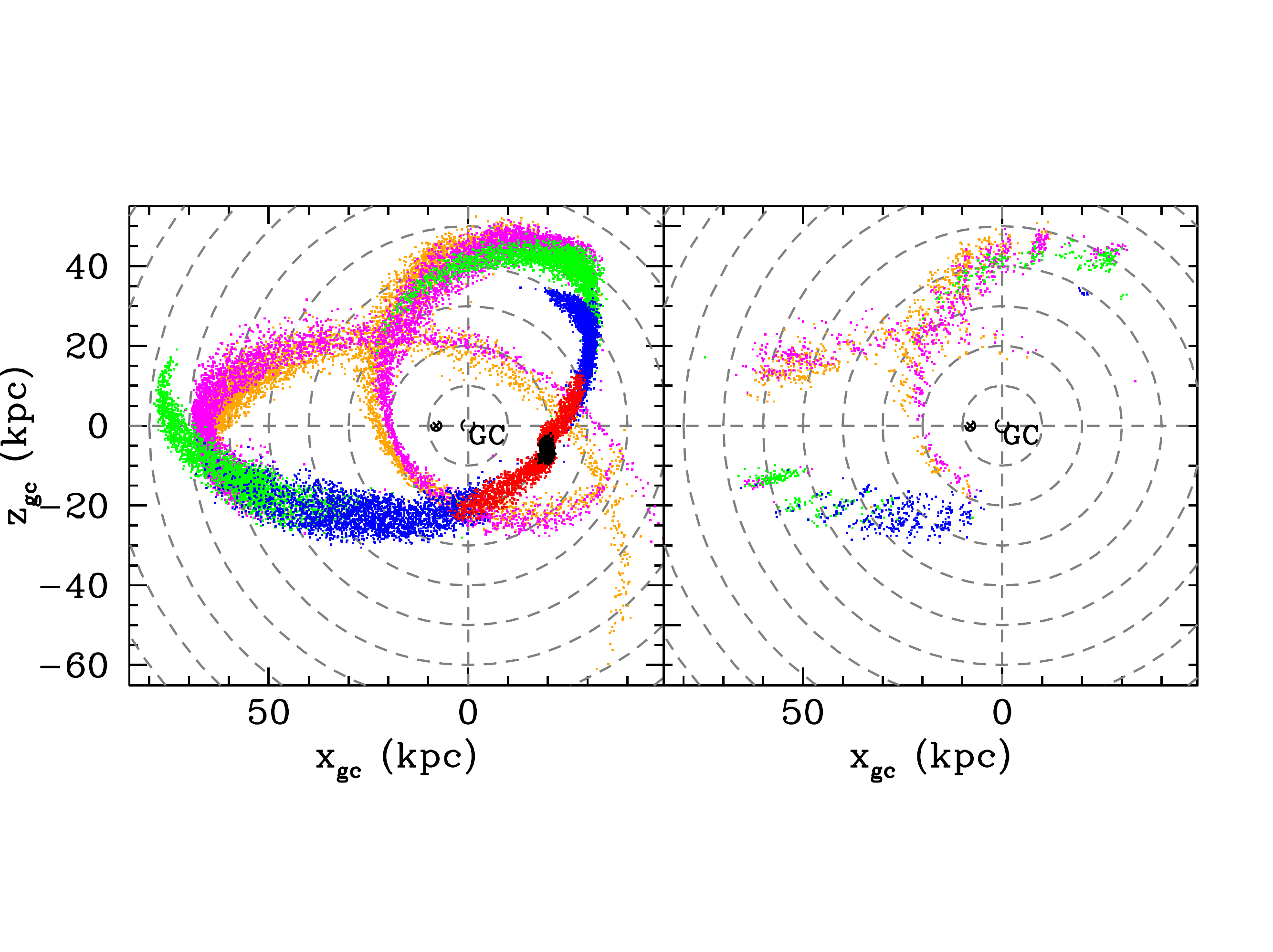} 
\end{center}
\caption{The LM10 model observed with SEGUE plate centers, shown in the $x_{gc}-z_{gc}$ plane, which is close to the Sgr orbital plane. In the left panel, each color represents stars lost on a different pericentric passage in the model (black shows bound stars; red, the most recent passage; blue, the next most recent passage; green, the second most recent; magenta, the third most recent; orange, the fourth most recent). The right panel shows the observed model, with the same colors. In both panels, dashed lines are every 10 kpc in $R_{gc}$. \label{modelplates}}
\end{figure}

\begin{figure}[t!]
\begin{center}
\includegraphics[scale=0.6]{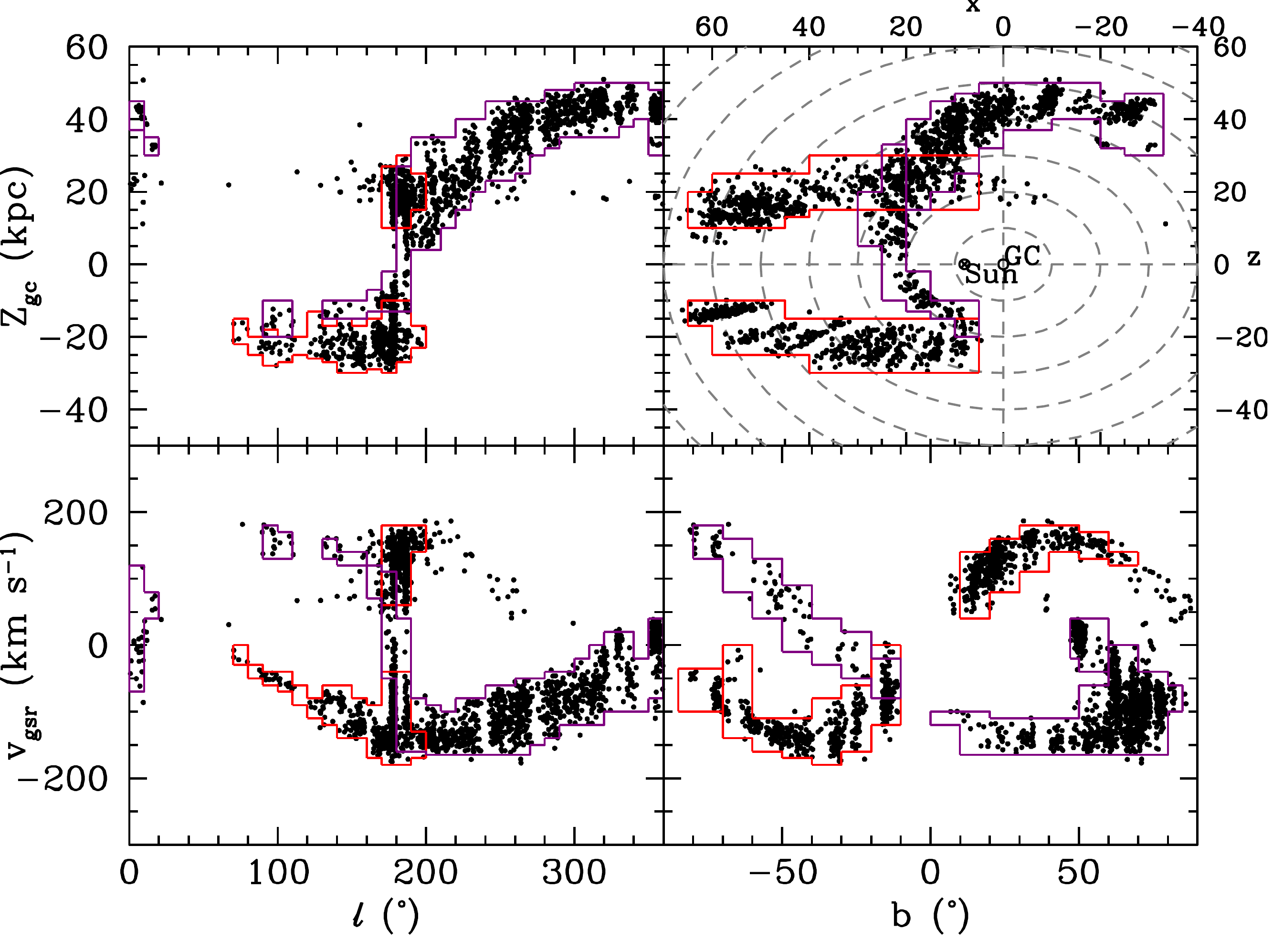} 
\end{center}
\caption{The regions defined as Sgr by use of the observed \citet{lm10} model, shown with the line-of-sight velocity (corrected for the solar and LSR motion) $v_{gsr}$, against Galactic longitude $l$ (lower left), Galactic latitude $b$ (lower right) and with longitude plotted against Galactocentric height from the plane $z_{gc}$ (upper left). The upper right panel shows the $x-z$ plane, which is close to Sgr's orbital plane. Dashed lines are every 10 kpc in $R_{gc}$. The observed model points are shown in black. The red boxes enclose streams identified as trailing in LM10, and the purple boxes enclose leading streams. \label{modelboxes}}
\end{figure}

Figure~\ref{groupscomsgr} shows groups that are classified as ``definitely Sgr," which appear to have a narrower velocity distribution than the \citetalias{lm10} model, as well as a slightly different distance distribution, where observed stars appear to be on average closer to the Sun in some regions (notably in Figure~\ref{groupscomsgr} along the northern leading stream, where the average $R_{gc}$ differs by as much as 10 kpc, and the width of the velocity distribution is 50 km s$^{-1}$ narrower in some places). These groups are also largely consistent with the positions of the Sgr streams using SDSS K/M-giants in \citet{yanny09b}, and using multiple target types by \citet{koposov12}. The \citetalias{lm10} model largely reproduces the observed distribution of Sgr stars in both 2MASS \citep{majewski03} and SDSS \citep{ruhland11}, pointing to a generally well constructed model. However, LM10 themselves acknowledge that their model does not produce features observed by SDSS (notably the bifurcation in the leading arm), so it is possible that other small deviations exist. Further observations are required to confirm Sgr membership and resolve these inconsistencies. 

One of our smaller groups (six members; marked in cyan in Figure~\ref{groups49}) has a large Galactocentric radius that is consistent with \citet{newbergsgr90}'s detection of potential Sgr debris 90 kpc from the Galactic center at $(l,b)=(190^\circ,30^\circ)$. Debris is also found in this location by \citet{ruhland11}, using a large sample of SDSS blue horizontal branch stars, as well as evidence of an extension to the trailing Sgr stream toward 90 kpc, for which we do not find FoF evidence. Further evidence for Sgr debris near this position was presented in \citet{drakesgr100kpc}. Using RR Lyrae stars, \citet{drakesgr100kpc} find a structure, which they call the Gemini stream, located near $(l,b)=(195^\circ,20^\circ)$ and extending to a distance of $\sim 100$ kpc. All of these structures are located in or near the Sgr plane, though their radial velocities are unmeasured. The distant trailing stream hypothesis is further supported by observations presented in \citet{belokurov14}, which found new Sgr trailing stream detections using photometry and kinematics in the Northern Galactic hemisphere. The results of \citet{koposov15} also support the distant trailing stream with spectroscopic observations of M giants in the area of the Galactic anticenter, consistent with the results of \citet{drakesgr100kpc} and \citet{belokurov14}.

It has been proposed that NGC2419 (with $(l,b)=(180^\circ,25^\circ)$ and $v_{gsr}=-14$ km s$^{-1}$) is associated with this structure \citep{newbergsgr90}, though the Gemini stream from \citet{drakesgr100kpc} is inconclusively linked to this cluster. Our FoF group has a position of $(l,b)=(187^\circ,15^\circ)$ and a $v_{gsr}$ of $-64$ km s$^{-1}$, making it possibly associated with either the Gemini stream or the \citet{newbergsgr90} structure. The \citetalias{lm10} model does not predict debris in this position. Streams at large Galactocentric radius are likely lost on early passages, and are particularly important to constrain both Sgr's accretion and the mass of the Milky Way.

\begin{figure}[t!]
\begin{center}
\includegraphics[scale=0.6]{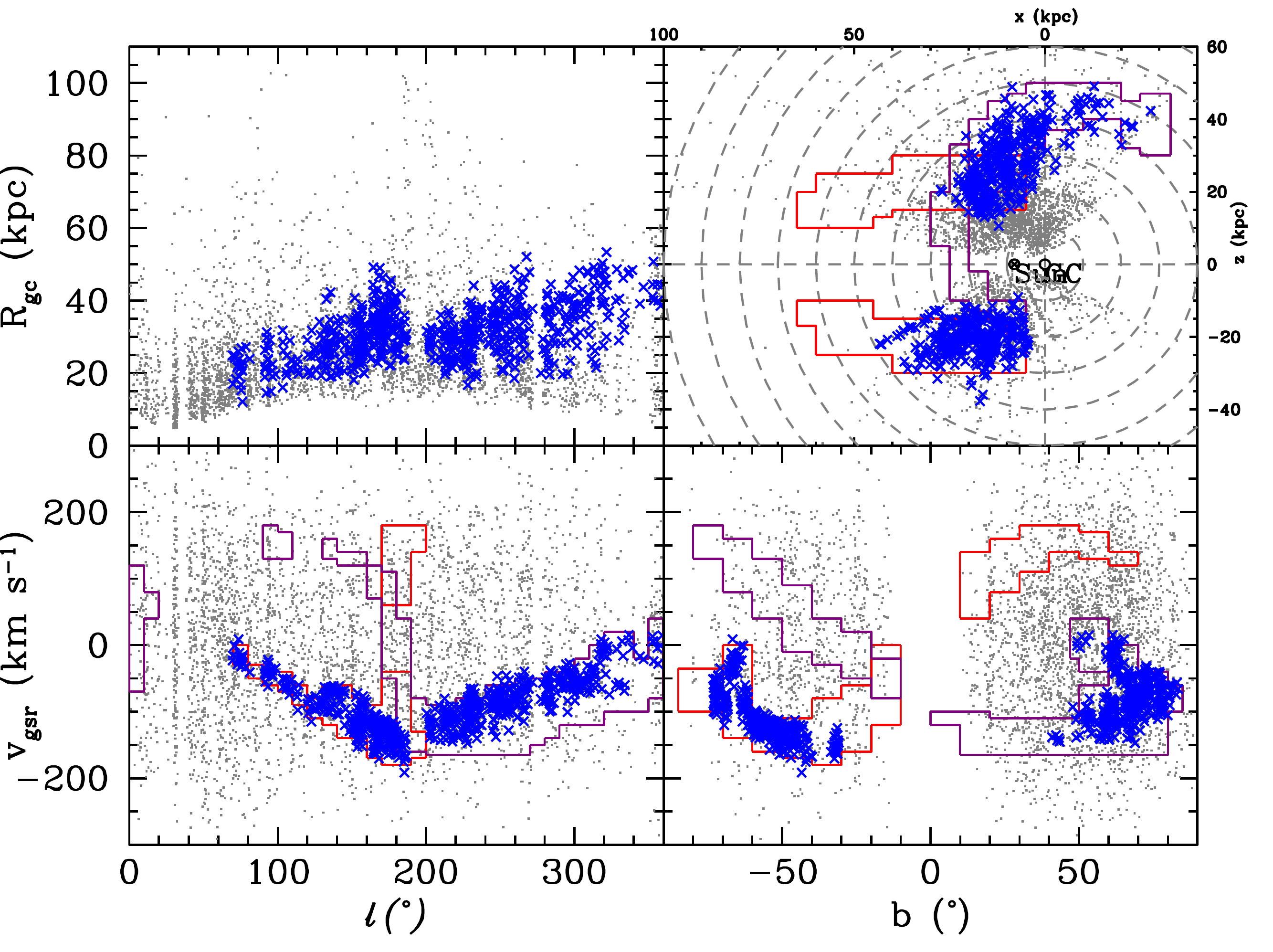}
\end{center}
\caption{All stars classified as part of groups `definitely' belonging to Sgr, shown in blue crosses. The panels are the same as those described in the caption for Figure~\ref{groups10}. This classification is made by using the method described in section \ref{sgrboxes}. The observed K giant substructure does not cover the extent of the LM10 model's velocity prediction for the leading stream between $l=220^\circ$ and $l=360^\circ$. \label{groupscomsgr}}
\end{figure}

\subsection{Attributing Groups to Other Halo Substructure}\label{coolstuff}
\begin{figure}[t!]
\begin{center}
\includegraphics[scale=0.6]{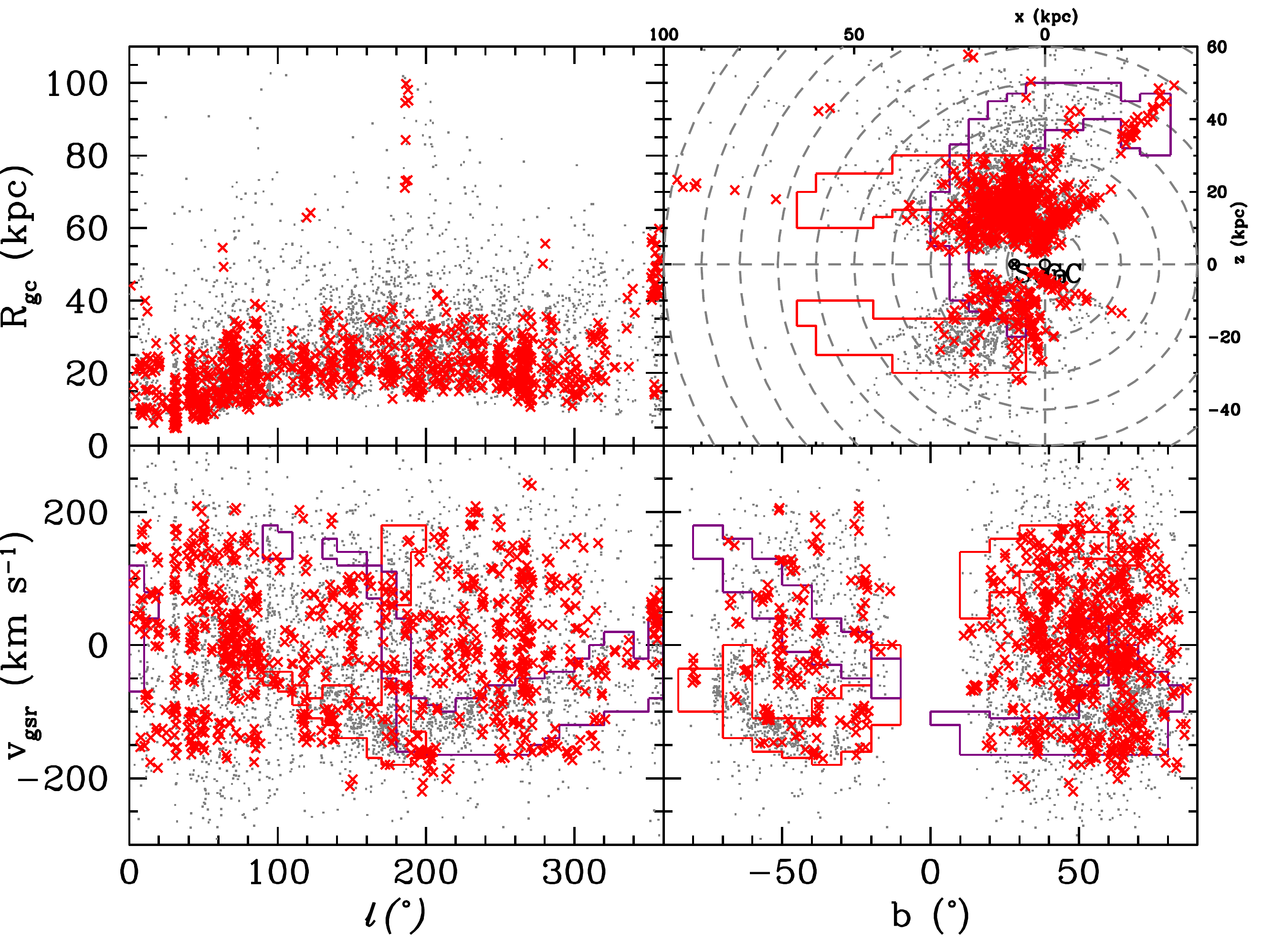}
\end{center}
\caption{All groups classified as `definitely not' belonging to Sgr. This classification is made by using the method described in section \ref{sgrboxes}. \label{groupscomnonsgr}}
\end{figure}
While Sgr is a significant component of the Milky Way halo, it is by no means the only visible substructure. Among the more interesting of other known substructures are the Orphan Stream \citep{belokurov}, the Virgo Overdensity \citep{juric08}, and Monoceros Ring \citep{belokurov}; the Grillmair-Dionatos stream \citep{gd06}; and the Cetus Polar stream \citep{newbergcetus}. Our group classification catalog contains additional classifications beyond those for general Sgr/non-Sgr groups. These classifications are designed to identify groups potentially associated with the Orphan Stream and Cetus Polar Stream. Our choice of FoF linking length is not well suited to detecting some substructure; we are unlikely to find a group that covers the extent of the Virgo Overdensity due to its large spatial size. The Grillmair-Dionatos stream is quite narrow and relatively close; a linking length chosen to find substructure at a given distance will create larger groups at a closer distance and ``wash out'' narrow streams. Finally, our criteria for removing the disk from our sample makes it difficult for us to detect the Monoceros ring. Our selection boxes are designed in a similar manner to those for Sgr, but instead of using a model, we use observational data of the streams from SDSS detections (\citet{newbergcetus}, \citet{newbergorph}). 

We identify several relatively large groups that are candidates for Orphan Stream and Cetus Polar Stream membership, with velocities and distances consistent with observations in \citet{newbergorph}, and \citet{newbergcetus} and \citet{koposov12}, respectively. These groups represent a small fraction of our overall sample, but are likely to be members of the Orphan and Cetus Polar streams. Aside from their matching kinematic data, the groups have [Fe/H] consistent with observations for each stream. Interestingly, both the Orphan and Cetus Polar streams are spatially coincident with Sgr in portions of the sky, so finding these groups is an illustration of the power of the FoF algorithm combined with 4-dimensional spatial and kinematic data. These groups are marked in Figure~\ref{groups49}.

We find one high latitude distant group that is spatially coincident with Sgr (group 397), and at roughly the same distance, but which has a strikingly different $v_{gsr}$ ($\Delta v \approx 200$ km s$^{-1}$). While this group has $(l,b)$ and $v_{gsr}$ consistent with observed substructure in Virgo \citep{newberg07}, it has a much greater distance. That this group may be associated with debris lost on an early pericentric passage of Sgr. In a future paper in this series (Z. Ma et al., in preparation), we further investigate this group.

Other groups listed in Table \ref{foftable} may belong to known substructures. Two groups with $l \approx 260^\circ$ and $b \approx 60^\circ$ (group 220 and group 211) have stars with $v_{gsr}$ consistent with the Virgo Stellar Stream \citep{duffau06}, but are over 10 degrees from the observed stream, and could be possible extensions to the Virgo Stellar Stream. We also detect a number of possible members of the Virgo Overdensity. In particular, groups 175 and 293 share the spatial and kinematic properties of the Virgo Overdensity \citep{bonaca12}. We will also further examine these stars, and in particular their [$\alpha$/Fe] properties, in Z. Ma et al. (in preparation).

Three groups with $l<45^\circ$  are located near the Galactic center with a large range of $v_{gsr}$. Their origin is unclear, but they could be associated with the Hercules-Aquila cloud \citep{belokurovha}. %It is possible that a number of groups listed in Table \ref{foftable} as `unlikely Sgr' are in fact associated with Sgr and were not captured by the box method described above, given that they fall along the Sgr streams in several plot quadrants in Figure~\ref{groups10}. 

\citet{grillmairhh} recently reported the discovery of two new halo streams, named Hermus and Hyllus. We find a group of five stars near the location of these streams (group 483). Further investigation of these streams and the members of the group are needed to conclusively determine their membership. 

\citet{koposov14} discovered a new metal poor stream in the ATLAS survey at a distance of 20 kpc, though it is mostly outside the SDSS footprint.  Finally, \citet{martin14} find a number of new streams in the vicinity of the Andromeda and Triangulum Galaxies using the PAndAS survey. Both of these locations are near the edge of our Milky Way disk star exclusion cuts, so we find no groups related to these streams. 

We provide the full catalog of stars in groups with 4 members or larger in Table \ref{groupcatalog}.

\subsection{FoF Discussion}
\subsubsection{False positives}

A number of false positive groups are expected because our choice of linking length does not change with $R_{gc}$. These false positive groups occur especially in the inner halo due to its higher density. In a smooth halo, we know that any identified groups are simply chance groupings of stars with similar positions, velocities, and distances. With a sample of stars, however, we expect there to be larger groups, consistent with the clustering of stars in substructure. For a data sample and a smooth halo model of equivalent size, we expect to find both a smaller number of total stars in groups and a smaller average group size in the smooth halo model than in the strucutred data sample.

% $\sim 600$ groups and
% $\sim 550$ groups and\begin{figure}[t!]

We have computed the expected number of groups in a smooth halo by generating ten model smooth halos, as described in section \ref{smoothdescription} above. We find that averaged over the ten halos, we expect to find $\sim 1600$ stars in groups with an average group size of 2.6, compared to $\sim 2200$ stars in groups in our K giant sample with an average group size of 4, leading to a higher average number of stars per group in the K giant sample by 1.4. These numbers are even more drastic when considering groups of three or more members, where there are 1550 stars in groups in the K giant sample, but only 800 stars in groups averaged over the ten smooth halo models, a factor of nearly two. We can conclude that a large fraction of our small groups (with a size of 3 members or less) are likely to be chance pairings and therefore false positives, but the fraction of these false positive groups decreases with group size.

\begin{figure}
\begin{center}
\includegraphics[scale=0.75]{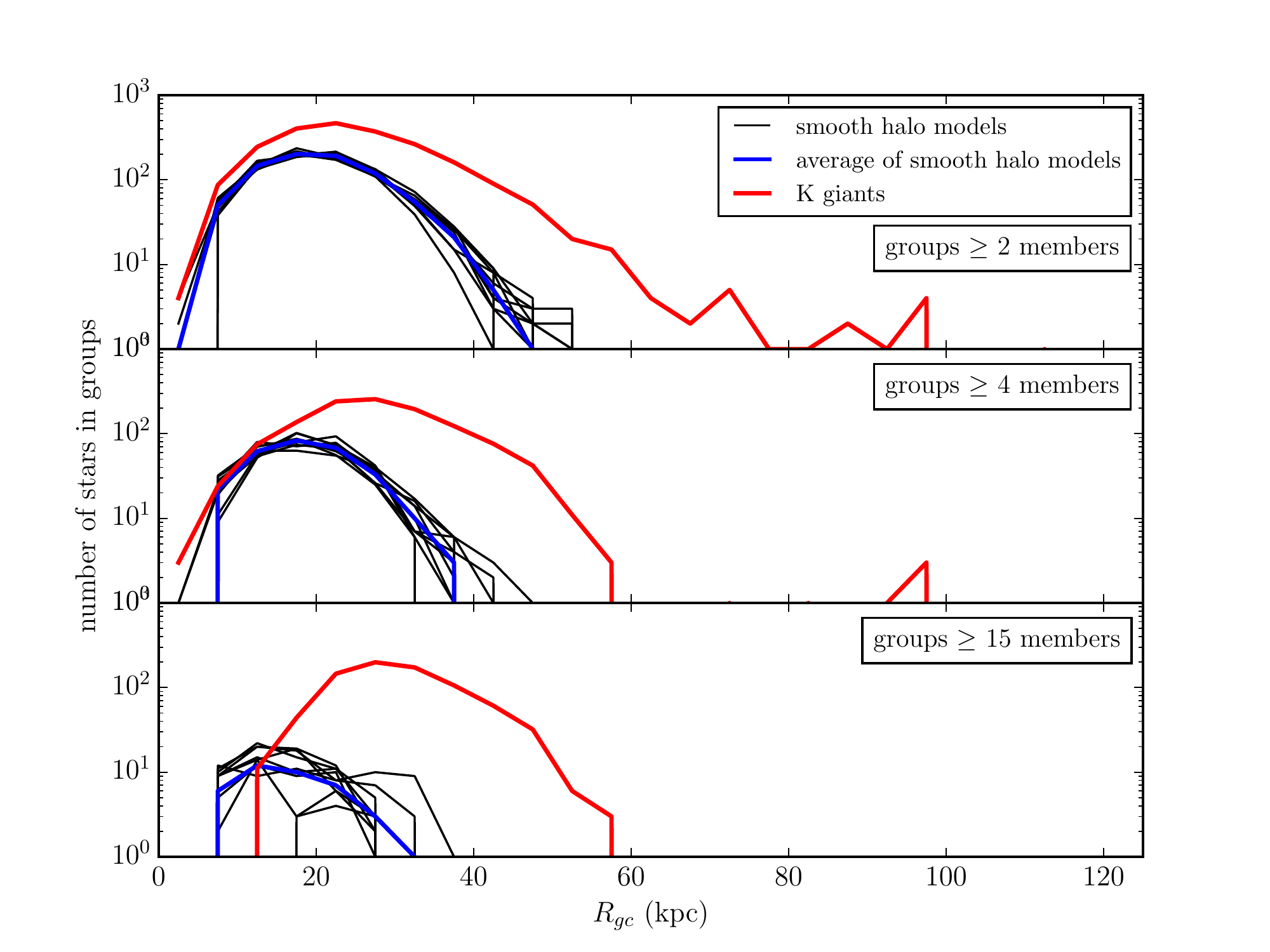}
\end{center}
\caption{Number of stars in groups as a function of Galactocentric radius, for both the K giant sample and for model smooth halos. Red lines show the number of stars in groups for the K giant sample, black lines show the number of stars in groups for each of ten individual smooth halo models. The blue line shows the average number of stars in groups for the smooth halo models. Top panel: All stars in groups of any size (two or more members). Middle panel: Stars in groups of four or more members. Bottom panel: Stars in groups of ten or more members. In all cases, there are more total groups in the K giant sample than on average in the smooth halo models. \label{falsepos}}
\end{figure}

In addition, we have computed the number of stars in groups as a function of Galactocentric radius for a number of group sizes. We show the results of this analysis in Figure \ref{falsepos}. In the top panel of Figure \ref{falsepos} we see that the number of stars in groups in the smooth halo models (blue line) is significant, but not as large as the number of stars in groups in the K giant sample at any Galactocentric radius, by at least a factor of two at most radii. There are almost no falsely grouped stars at radii greater than 40 kpc. The trend continues for groups of larger size as well, with groups of four members or more and groups of ten members or more showing significantly more stars in groups at all but the most nearby Galactocentric radii.

It is also possible for our method to miss cold substructure \citep[e.g., the Grillmair-Dionatos stream;][]{gd06} in the higher density inner regions of the halo, because our choice of linking length are designed to identfy cold substructure in more distant regions of the halo. In effect, because of the higher stellar density, more pairs per star will be found, increasing a group's size and the range of its values in the 4-dimensional parameter space, and ``washing out'' the cold stream. 
%Thus we would expect nearby structures picked up by FoF to be broad.

\subsubsection{Group membership trends}
Figure~\ref{groupsizetype} shows a histogram of the number of groups with a given group size and their classifications. While the largest groups (larger than 10 members) are predominantly classified as Sgr, the smallest groups (10 or fewer members) are mostly unlikely to be Sgr, and only 13\% of all groups are likely to be members of any previously known substructure. 

We can, therefore, give limits on the fraction of halo stars residing in substructure. At least 50\% of K giants in our sample show no detection of substructure. Additionally, at least 13\% of stars in our sample are members of Sgr, and roughly 1\% of stars are members of other known substructure. The remaining $\sim 36$\% of stars in the sample comprise previously undiscovered substructure.

\begin{figure}[t!]
\begin{center}
\includegraphics[scale=0.6]{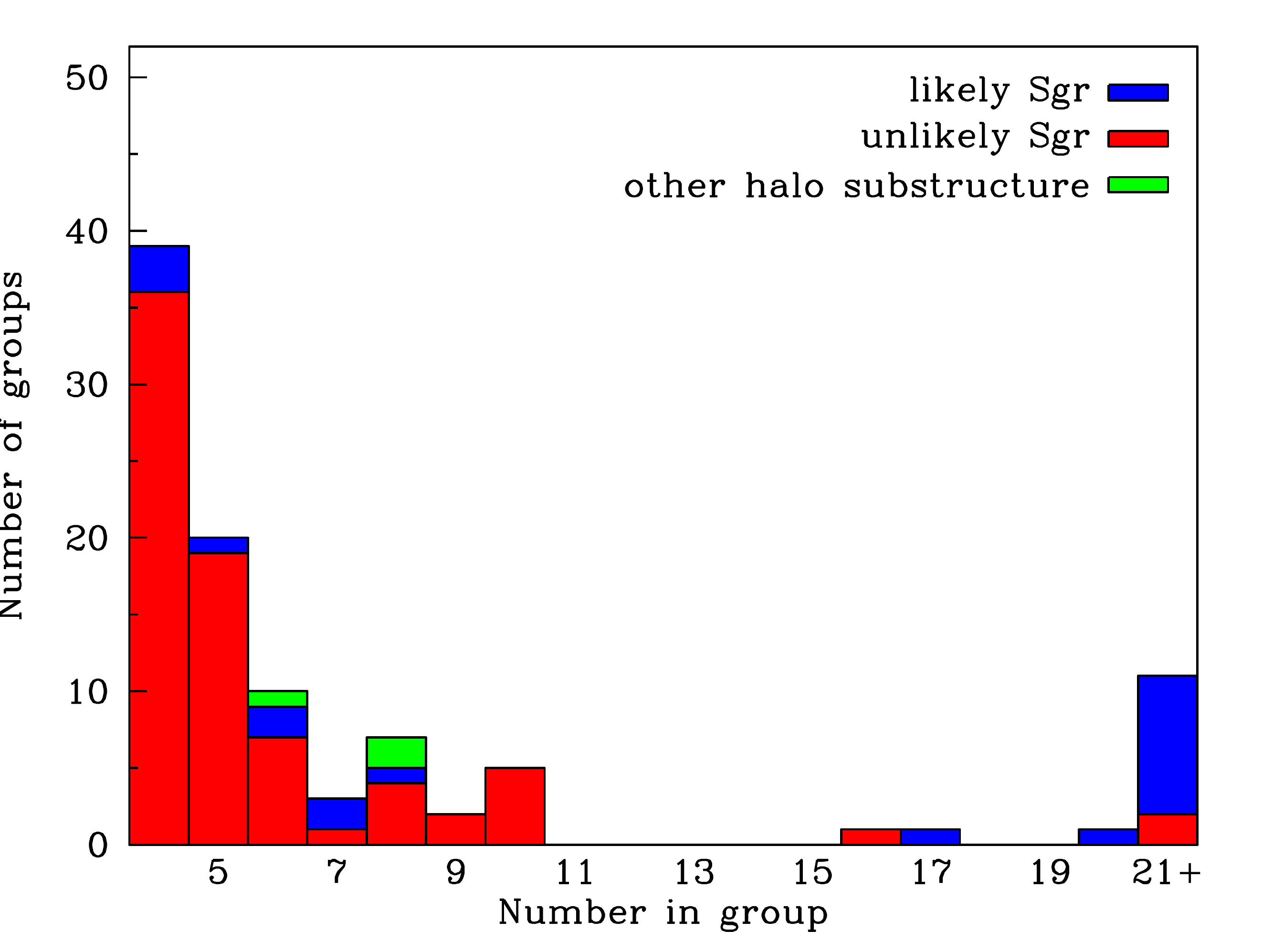}
\end{center}
\caption{Number of groups at a given group size, classified by their associated structure. Blue are groups within Sgr regions, green and red are groups within non-Sgr regions. The rightmost bin is the combination of all groups larger than 20 members: the largest group contains 255 stars. For groups larger than 20 members, 80\% of groups are associated with Sgr. Overall, approximately 700 stars are associated with Sgr. \label{groupsizetype}}
\end{figure}

Figure~\ref{sgrwrgc} shows the fraction of stars in groups that belong to Sgr groups as a function of $R_{gc}$ in different samples of K giants. For the full K giant sample, we see that about 1/3 of stars in groups belong to a Sgr group, and at least 50\% of stars with $R_{gc} >$ 30 kpc are Sgr members. This trend also helps to explain our earlier finding of an increase in substructure with $R_{gc}$, as Sgr is a greater percentage of the total stellar population as $R_{gc}$ increases. The Figure also shows four metallicity selected samples. It is clear from this diagram that in the more metal poor samples, there is a smaller fraction of Sgr stars than in the more metal rich samples. In fact, for some $R_{gc}$ bins, Sgr represents more than 75\% of the stars. Since the Sgr stream is by far the dominant structure found in our analysis, this result readily explains the trend of increasing substructure with metallicity, and much of the inner/outer halo differences found in Section \ref{4dresults}. This conclusion also suggests an alternate explanation for the lower
level of substructure found by X11 and \citet{cooper11} in BHB stars compared to the simulations: not age, but metallicity. Metallicity plays into substructure detection via the mass-metallicity
relation: more massive satellites contain more metal-rich stars and
are also easier to detect in pencil-beam surveys such as SEGUE because
of the higher stellar density in their streams. Stars from low-mass
satellites will be metal-poor and hard to detect because of their
lower stellar density and the wide spacing of the SEGUE fields. 
 
\begin{figure}[t!]
\begin{center}
\includegraphics[scale=0.6]{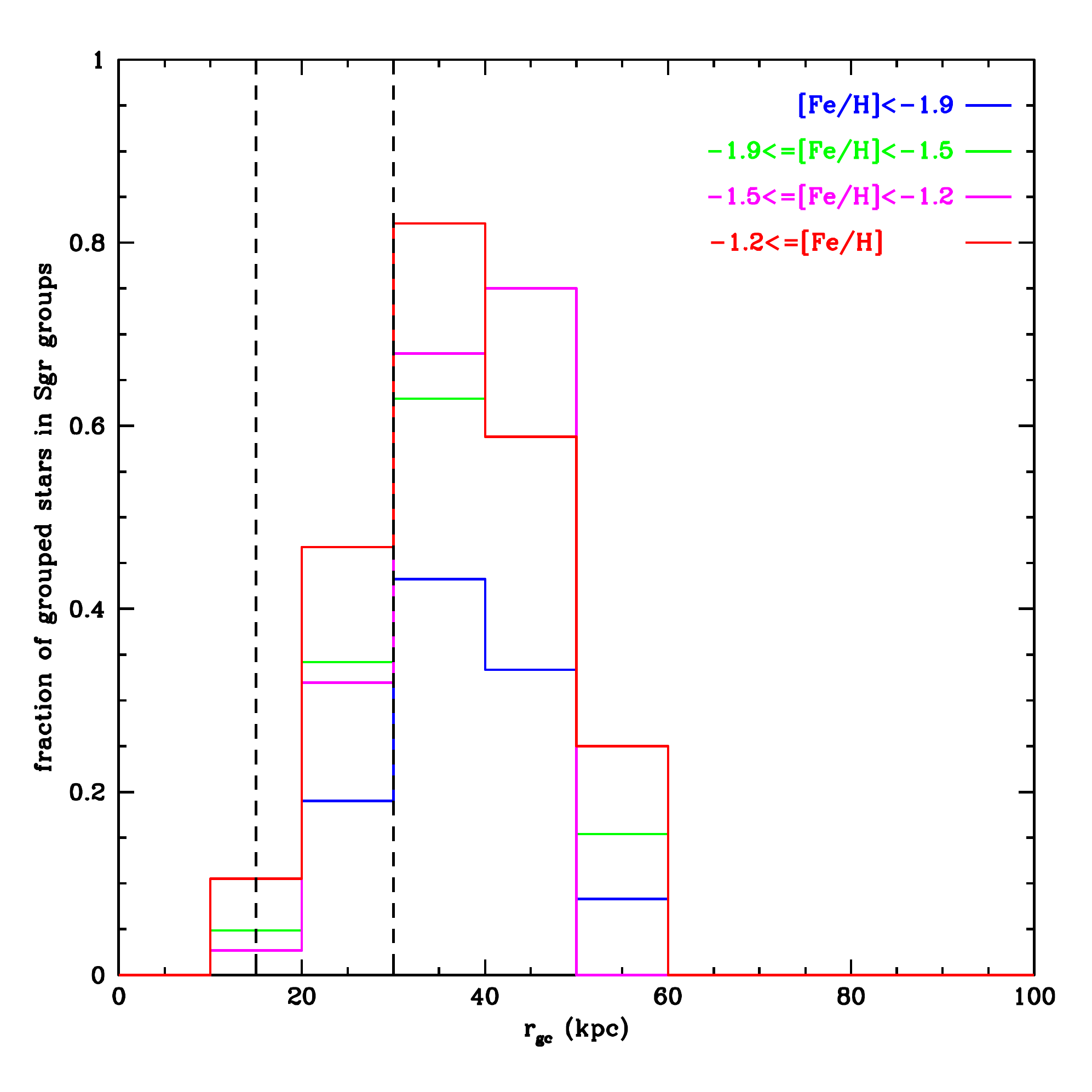}
\end{center}
\caption{Fraction of stars in groups associated with Sgr as a function of Galactocentric radius. The colored lines show different metallicity samples discussed in Section \ref{pairsmetal}. Dashed vertical lines indicate the distance ranges over which we divide our sample into $R_{gc}$ bins in Figure~\ref{kgnrf}. 
Approximately 1/3 of stars in groups are associated with Sgr, and the most metal rich sample is largely composed of Sgr stars. \label{sgrwrgc}}
\end{figure}

\section{Conclusions}

We use a sample of 4568 halo K giants with distances up to 125 kpc to
measure substructure in the halo using a metric sensitive to both
spatial and kinematical substructure. This sample complements and
extends the work of X11 and \citet{cooper11} on substructure in large
samples of BHB stars because K giants are not restricted to old, metal
poor populations. 

Outer halo K giants show more substructure than inner halo ones, in agreement with
the results of X11 and \citet{cooper11}. In addition, we find that the most metal-rich K
giants in our sample (with \fe\ $\ge -1.2$) show the most substructure
of all the K giants. In addition, when directly comparing the amount of substructure
in the BHB sample from X11 and the K giants presented in this work, we find no significant difference
between the two samples when selecting stars over equivalent distance ranges. Further work is needed to 
fully understand the contribution of BHBs and K giants to the Galactic stellar halo.

Since Sgr stream stars are on average more metal rich than the rest of
the halo and the Sgr stream is not found in the inner halo, we
investigated the possibility that the Sgr stream is responsible for
both trends in substructure (metallicity and distance). We find that
for stars with \fe\ greater than --1.9, most of the groups with
$R_{gc}$ greater than 30 kpc belong to the Sgr stream, and conclude
that in the case of the Milky Way, the Sgr stream is responsible for
the increase of substructure with both metallicity and distance.

\acknowledgments
We thank Mario Mateo and Eric Bell for enlightening discussions.  HLM, CMR and PH thank the Aspen Center for Physics (and NSF grant PHYS-1066293) for hospitality during the writing and editing of this paper.

WFJ, HLM, and ZM acknowledge support from grants AST-1009886 and AST-121989 to HLM.

ES gratefully acknowledges the Canadian Institute for Advanced Research (CIFAR) for financial support.

XX acknowledges the Alexander von Humboldt foundation for a fellowship, the support of the Max-Planck-Institute for Astronomy, and the support by the National Natural Science Foundation of China under grant Nos. 11103031, 11233004, 11390371.

TCB acknowledges partial support for this work from grant PHY 14-30152; Physics Frontier Center/JINA Center for the Evolution of the Elements (JINA-CEE), awarded by the US National Science Foundation.

YSL acknowledges support provided by the National Research Foundation of Korea to the Center for Galaxy Evolution Research (No. 2010-0027910) and the Basic Science Research Program through the National Research Foundation of Korea (NRF) funded by the Ministry of Science, ICT \& Future Planning (NRF-015R1C1A1A02036658).

Funding for SDSS-III has been provided by the Alfred P. Sloan Foundation, the Participating Institutions, the National Science Foundation, and the U.S. Department of Energy Office of Science. The SDSS-III web site is http://www.sdss3.org/.

SDSS-III is managed by the Astrophysical Research Consortium for the Participating Institutions of the SDSS-III Collaboration including the University of Arizona, the Brazilian Participation Group, Brookhaven National Laboratory, University of Cambridge, Carnegie Mellon University, University of Florida, the French Participation Group, the German Participation Group, Harvard University, the Instituto de Astrofisica de Canarias, the Michigan State/Notre Dame/JINA Participation Group, Johns Hopkins University, Lawrence Berkeley National Laboratory, Max Planck Institute for Astrophysics, Max Planck Institute for Extraterrestrial Physics, New Mexico State University, New York University, Ohio State University, Pennsylvania State University, University of Portsmouth, Princeton University, the Spanish Participation Group, University of Tokyo, University of Utah, Vanderbilt University, University of Virginia, University of Washington, and Yale University.

\begin{deluxetable}{lrrrrrrr}
\tablewidth{0pt}
\tablecaption{Selected\tablenotemark{a} FoF groups found at linking length $4\delta=0.03$ \label{foftable}}
\tablehead{\colhead{Group ID} & \colhead{Members} & \colhead{$l$\tablenotemark{b}} & \colhead{$b$\tablenotemark{b} }& \colhead{$v_{gsr}$\tablenotemark{b}} & \colhead{$R_{gc}$\tablenotemark{b}}& \colhead{[Fe/H]\tablenotemark{b}} & \colhead{Notes}\\ \colhead{} & \colhead{~} & \colhead{deg} & \colhead{deg} & \colhead{km s$^{-1}$} & \colhead{kpc} & \colhead{dex} & \colhead{~}}
\startdata
3 & 212 &  251.48 &   68.18 &  -85.77 & 31.98 & -1.39 & Sgr \\
11 & 174 &  162.10 &  -53.62 & -123.26 & 30.55 & -1.19 & Sgr \\
41 & 59 &  143.72 &  -70.97 &  -80.86 & 27.38 & -1.30 & Sgr \\
28 & 54 &  300.51 &   74.36 &  -56.40 & 36.15 & -1.34 & Sgr \\
40 & 50 &   88.08 &  -67.66 &  -34.97 & 22.65 & -1.21 & Sgr \\
210 & 34 &  208.45 &   60.65 & -115.48 & 27.93 & -1.48 & Sgr \\
369 & 27 &  217.83 &   52.33 & -109.82 & 27.68 & -1.18 & Sgr \\
9 & 24 &  340.01 &   49.86 &   49.85 & 46.70 & -1.49 & unlikely Sgr \\
5 & 23 &  323.22 &   62.38 &   -7.61 & 43.89 & -1.18 & Sgr \\
110 & 21 &  153.98 &  -51.47 & -133.08 & 31.16 & -1.24 & Sgr \\
107 & 20 &  209.58 &   43.17 &  -98.98 & 25.61 & -1.34 & likely Sgr \\
376 & 17 &  166.29 &  -31.85 & -149.72 & 39.24 & -1.26 & Sgr \\
220 & 16 &  264.95 &   56.62 &   24.56 & 20.24 & -1.79 & unlikely Sgr \\
332 & 10 &   71.63 &  -50.49 &   25.61 & 27.75 & -1.41 & not Sgr \\
124 & 10 &  135.13 &  -54.99 & -105.92 & 25.55 & -1.39 & unlikely Sgr \\
211 & 10 &  172.49 &   65.68 &   64.95 & 20.36 & -1.28 & unlikely Sgr \\
71 & 10 &   42.12 &   35.69 &   18.20 & 20.34 & -1.59 & possible Hercules-Aquila Cloud\tablenotemark{c} \\
222 & 10 &  258.33 &   55.49 &  -35.96 & 17.24 & -1.39 & unlikely Sgr \\
26 & 9 &   43.57 &   35.84 & -148.32 & 12.81 & -1.55 & possible Hercules-Aquila Cloud\tablenotemark{c} \\
175 & 8 &  288.81 &   60.93 &  103.14 & 17.72 & -1.34 & possible Virgo Overdensity \\
32 & 8 &  145.52 &  -46.47 &  -54.29 & 35.85 & -2.13 & Cetus Polar Stream\\
408 & 8 &  183.93 &   48.88 &  130.70 & 49.07 & -2.11 & Orphan Stream\\
275 & 7 &  328.32 &   82.11 &  -63.96 & 36.37 & -1.34 & Sgr \\
261 & 6 &   41.88 &   78.69 & -101.72 & 12.14 & -1.57 & possible Hercules-Aquila Cloud\tablenotemark{c} \\
388 & 6 &  187.08 &   14.80 &  -63.43 & 90.80 & -1.34 & possible distant Sgr\tablenotemark{d} \\
224 & 6 &  138.39 &  -71.76 &  -21.53 & 42.94 & -2.19 & Cetus Polar Stream\\
8 & 6 &  355.33 &   51.05 &    5.53 & 43.92 & -1.82 & Sgr \\ 
397 & 5 &  256.07 &   69.40 &  148.04 & 20.05 & -1.22 & possible older Sgr\tablenotemark{e} \\
483 & 5 &   77.21 &   45.34 &   -4.32 & 23.80 & -1.66 & possible Hermus/Hyllus\tablenotemark{f} \\
93 & 4 &  304.25 &   73.51 &  153.35 & 15.10 & -1.59 & possible Virgo Overdensity \\\enddata
\tablenotetext{a}{
~Larger than 10 members \emph{or} identified as notable structure
}
\tablenotetext{b}{
~Mean value for all stars in group
}
\tablenotetext{c}{
~see \citet{belokurovha}; well-matched in $l$, $b$, $r_gc$, [Fe/H], but not $v_{gsr}$
}
\tablenotetext{d}{
~see \citet{newbergsgr90}
}
\tablenotetext{e}{
~denoted as SgrP in \citet{paper4}
}
\tablenotetext{f}{
~see \citet{grillmairhh}
}
\end{deluxetable}

\begin{deluxetable}{lrrrrrrrrrrrrrrr}
  \rotate
\tablewidth{0pt}
\tablecaption{Group catalog: K giants sorted by group for groups with 4 or more members \label{groupcatalog}}
\tablehead{\colhead{Plate} & \colhead{MJD} & \colhead{Fiber} & \colhead{$l$} & \colhead{$b$} & \colhead{$v_{gsr}$} & \colhead{$v_{err}$} & \colhead{$d$} & \colhead{$d_{err}$} & \colhead{[Fe/H]} & \colhead{$g_{01}$} & \colhead{$M_r$} & \colhead{$r_{gc}$} & \colhead{Target} & \colhead{Group} & \colhead{Group }\\ \colhead{~} & \colhead{~} & \colhead{~} & \colhead{deg} & \colhead{deg} & \colhead{km s$^{-1}$} & \colhead{km s$^{-1}$} & \colhead{kpc}  &\colhead{kpc} & \colhead{dex} & \colhead{mag} & \colhead{mag} & \colhead{kpc} & \colhead{type\tablenotemark{a}}  &\colhead{ID} & \colhead{flag\tablenotemark{b}}}
\startdata
  519 & 52283  &  62 & 289.4253  & 64.0776  &  -48.7  &    1.4   &  37.2   &   4.2  &  -0.81   & 17.027  &  -1.98  & 36.862   &3  & 3  & 6\\
  2413 & 54169 &  215 & 246.7213 & 60.4712 &   -89.2 &     2.6  &   41.5  &    6.5 &   -1.02  & 18.665 &   -0.13 &  43.808  & 1 &  3 &  6\\
  2509 & 54180 &  597 & 237.3916 & 74.4764 &  -105.9 &     1.6  &   35.4  &    4.2 &   -1.22  & 17.310 &   -1.30 &  37.449  & 2 &  3 &  6\\
  2857 & 54453 &    5 & 231.0205 & 67.2724 &  -101.0 &     1.5  &   21.8  &    2.4 &   -1.52  & 15.808 &   -1.77 &  24.995  & 4 &  3 &  6\\
  2857 & 54453 &  177 & 228.0546 & 66.8574 &  -112.3 &     2.7  &   17.2  &    2.5 &   -2.52  & 16.289 &   -0.50 &  20.832  & 1 &  3 &  6\\
\enddata
\tablecomments{Table \ref{groupcatalog} is published in its entirety in the electronic edition of <>. A portion is shown here for guidance regarding its form and content.}
\tablenotetext{a}{
~1 -- l-color K giants; 2 -- proper motion K giants; 3 -- red K giants; 4 -- serendipitous K giants.
}
\tablenotetext{b}{
~3 -- not Sgr; 4 -- unlikely Sgr; 5 -- possible Sgr; 6 -- definitely Sgr; 7 -- Orphan Stream; 8 -- Virgo Overdensity; 9 -- Cetus Polar Stream.
}
\end{deluxetable}

\clearpage

\end{document}